\documentclass{aa}

\usepackage{graphicx}

\usepackage{natbib}
\usepackage{txfonts}

\begin{document}

\title{The chemical evolution of a Milky Way-like galaxy: the importance of a cosmologically motivated infall law} 

\titlerunning{A cosmological infall law for the Milky Way}

\author {E. Colavitti\inst{1,}\thanks{email to: colavitti@oats.inaf.it}
\and  F. Matteucci\inst{1,2}
\and  G. Murante\inst{3}}

\authorrunning{E. Colavitti et al.}

\institute{Dipartimento di Astronomia, Universit\'a di Trieste,  Via G. B. Tiepolo 11, I-34143 Trieste (TS), Italy 
\and  I.N.A.F. Osservatorio Astronomico di Trieste, Via G. B. Tiepolo 11, I-34143 Trieste (TS), Italy
\and  I.N.A.F. Osservatorio Astronomico di Torino, Strada Osservatorio 20, I-10025 Pino Torinese (TO), Italy}

\date{Received xxxx / Accepted xxxx}

\abstract{}
{We aim at finding a cosmologically motivated infall law to understand if the $\Lambda$CDM cosmology can reproduce the main chemical characteristics of a Milky Way-like spiral galaxy.}
{In this work we test several different gas
infall laws, starting from that suggested in the two-infall model
for the chemical evolution of the Milky Way by Chiappini et al., but focusing
on laws derived from cosmological simulations which
follows a concordance $\Lambda$CDM cosmology. By means of a detailed
chemical evolution model for the solar vicinity, 
we study the effects of
the different gas infall laws on the abundance patterns and the G-dwarf metallicity distribution.} 
{The cosmological gas infall law,
derived from dark matter halos having properties compatible with the
formation of a disk galaxy like the Milky Way, and assuming that the
baryons assemble like dark matter, resembles the infall law suggested
by the two-infall model. In particular, it predicts two main gas
accretion episodes. 
Minor infall episodes are predicted to
have followed the second main one but they are of small significance
compared to the previous two. By means of this cosmologically motivated
infall law, we study the star
formation rate, the SNIa and SNII rate, the total amount of gas and
stars in the solar neighbourhood and the behaviour of several chemical abundances.
We find that the results of the
two-infall model are fully compatible with the evolution of the Milky
Way with cosmological accretion laws.
We derive that the timescale
for the formation of the stellar halo and the thick disk must 
have not been longer than 2 Gyr, whereas the disk in the solar vicinity 
assembled on a much longer timescale ($\sim$ 6 Gyr).} 
{A gas assembly history derived from a DM halo,
compatible with the formation of a late-type galaxy from the
morphological point of view, can produce chemical
properties in agreement with the available observations.}{}


\maketitle

\section{Introduction}
In many models of the chemical evolution of the Milky Way gas infall has been invoked to explain the formation of the Galactic disk (e.g. Chiosi (1980), Matteucci \& Fran\c cois (1989); Lacey \& Fall (1985); Chiappini et al. (1997); Boissier \& Prantzos (2000) among others).  Originally, the gas infall was introduced as a possible solution to the G-dwarf problem (Pagel 1989). In general terms the gas infall rate gives the law for the assembly of baryons in a galaxy. However, in the majority of the chemical evolution papers existing in the literature, the gas infall law has been treated as a free parameter with no connection to a galaxy's cosmological context. In other words, in most of the cases the assumed infall law is independent of the details of the galactic dark matter (DM) halo's assembly which, instead, should have a dominant effect on it.
On the other hand, the infall law is clearly  very important in determining the main characteristics of a galaxy. 
In this paper we aim at studying the infall law which descends directly from 
the DM halo and its assembly. 

In this way, we will have an infall law for the gas which is related to cosmology and does not contain free parameters. Once achieved that, we will test this cosmological infall law in a detailed model of chemical evolution of the Milky Way which follows the evolution of many chemical species by taking into account the stellar lifetimes, detailed nucleosynthesis prescriptions and supernova (type II, Ib/c and Ia) rates. Several authors have tried before us to build a model for the evolution of the disk galaxies in a cosmological context, but none of these considered the chemical evolution in such a detail as our model.  
Chemo-dynamical models for the Milky Way were proposed by Theis, Burkert \& Hensler (1992), where the evolution of massive spherical galaxies was calculated by a multi-component hydrodynamical approach but with no cosmological context.
In Raiteri, Villata \& Navarro (1996), instead,  N-body/hydrodynamical cosmological simulations were used to investigate the chemical evolution of the Galaxy by assuming that it formed by the collapse of a rotating cloud of gas and dark matter. However, their chemical analysis, although detailed, was limited to only oxygen and iron. Another important chemodynamical paper appeared
in Samland, Hensler \& Theis (1997), in which they presented their two-dimensional chemodynamical code CoDEx. Their model contains nucleosynthesis from supernovae of type I and II and some chemical evolution, but no cosmological context was assumed.
More recently, Abadi et al. (2003) presented simulations of galaxy formation in a $\Lambda$ cold dark matter universe ($\Lambda$CDM) and studied the dynamical and photometric properties of disk galaxies, but no chemical evolution was included.

Robertson et al. (2005) adopted the hierarchical scenario for galaxy formation
to see if in this context they could reproduce the rich data set of stellar
abundances in the galactic halo and Local Group dwarf galaxies.
They used an analytical expression for the growth of DM halos in a
$\Lambda$CDM cosmology. Their baryonic infall law is proportional to the DM one.
The hierarchical formation
scenario, when applied to the stellar halo of the Milky Way, suggests that it
formed through accretion and disruption of dwarf galaxies.  They concluded
that the majority of stars in the stellar halo were formed within a relatively
massive dwarf irregular sized dark matter halos, which were accreted and
disaggregated $\sim 10$ Gyr ago. In their scenario, these systems had rapid star
formation histories and were enriched primarily by supernovae (SNe) of type
II. They also suggested that the still existing dwarf irregular galaxies
formed stars more gradually and they underwent both SNIa and II enrichment. On
the other hand, dwarf spheroidal galaxies should be systems where the
abundances are determined by galactic winds. In summary, the paper dealt
mostly with the comparison between the [$\alpha$/Fe] ratios in the galactic
halo and dwarf galaxies.

Also Naab \& Ostriker (2006) studied the metallicity and photometric evolution of a generic disk galaxy, by assuming that it forms through mergers of dark matter halos. 
They took a point of view similar to that of the present paper: in particular, they derived a cosmological infall law and concluded that the infall rate should have been almost constant during the lifetime of the disk. 
No detailed chemical evolution was followed and no consideration was given to the formation of the stellar halo.

Finally, another paper dealing with chemical evolution in a cosmological context is that by Nagashima \& Okamoto (2006). 
The authors investigated the chemical evolution in Milky Way-like galaxies based on the CDM model in which cosmic structures form via hierarchical merging. They adopted a semi-analytical model for galaxy formation where the chemical enrichment due to both SNeIa and SNeII was considered.
They suggested that the so-called G-dwarf metallicity problem can be fully resolved  by the hierarchical formation of galaxies. In fact, the infall term introduced by the traditional monolithic collapse models to solve the G-dwarf problem can be explained by some physical processes such as injection of gas and metals into hot gas due to SNe.
The model, however, was not tested on large number of chemical elements but was limited to the [O/Fe] vs. [Fe/H] plot.

In this paper we will first study the effect of different gas infall laws
taken from the literature and compare the chemical results with those of
Fran\c cois et al.'s (2004) model, which is based on the two-infall model of
Chiappini et al. (1997). In the two-infall model it is assumed that the halo
and the thick disk formed by means of a first infall episode on a timescale
not longer than 2 Gyr, whereas the thin disk should have formed by means of an
independent second infall episode lasting much longer. In particular, the
timescale for the formation of the solar vicinity was 7 Gyr, as suggested by
the G-dwarf metallicity distribution, while the internal parts of the thin
disk formed faster and the outermost regions are still forming now. This
scenario has proven to be very successful in reproducing the majority of the
properties of the solar vicinity and the whole disk and it was adopted by the
majority of the chemical evolution models of the Milky Way. In Fran\c cois et
al. (2004) the evolution of 35 chemical species including C, N, O,
$\alpha$-elements (Ne, Mg, Si, Ca, Ti), Fe-peak elements plus light
elements such as D, He and $^{7}$Li is followed in detail.  In this paper we
run a cosmological simulation to find a suitable Dark Matter (DM) halo for a
Milky Way-like galaxy by adopting GADGET2 (Springel 2005) and to obtain the infall law for
the gas. In particular, we derive the law for the accretion of the DM
halo by assuming that the same law is followed by the assembling
baryons. Once obtained, this law is tested in the chemical evolution
model to see if it is consistent with the two-infall or other scenarios. In
order to do that we calculate in detail the evolution of the abundances
of several chemical elements, the SN rates and all the physical quantities
relevant to the evolution of the solar vicinity. Therefore, 
we start from a different approach relative to all the previous hierarchical
models for the formation of the Milky Way (but see Sommer-Larsen et al. 1999).
The reason for considering only gas accretion and not dwarf galaxies, as in other papers, is suggested by the different chemical histories observed in dwarf galaxies relative to the Milky Way (e.g. Lanfranchi \& Matteucci, 2004).

The paper is organized as follows: in section 2 we show the nucleosynthesis prescriptions adopted. Section 3 presents a brief description of the model by Chiappini et al. (1997). In section 4 we describe the cosmological simulation, done using the simulator Gadget2. Section 5 describes the adopted infall laws. In section 6 we present the results obtained, comparing the models predictions with the observed properties. Finally section 7 presents the conclusions.

\section{Nucleosynthesis prescriptions}

One of the most important ingredients for chemical evolution models is represented by the nucleosynthesis 
prescriptions and consequently by the stellar yields.

The single stars in the mass range $0.8 \; M_{\odot} \; \leq \; M \; \leq \; 8 \; M_{\odot}$ (low and intermediate-mass stars) contribute to the Galactic enrichment through planetary nebula ejection and quiescent mass loss. They enrich the interstellar medium mainly in He, C,  N and heavy s-process elements (e.g. Cescutti et al. 2006). 
We adopt here the stellar yields for low and intermediate mass stars of van den Hoek \& Groenewegen (1997) computed as functions of stellar metallicity, their case with variable mass loss.
These stars are also the progenitor of Type Ia supernovae (SNe), if they are in binary systems, which originate from carbon deflagration of C-O white dwarfs. We adopt in this paper the single-degenerate progenitor scenario (Whelan \& Iben, 1973; Han \& Podsiadlowski 2004). Type Ia SNe contribute a substantial amount of Fe ($\sim \; 0.6 \; M_{\odot}$ per event) and Fe-peak elements as well as non negligible quantities of Si and S. They also produce other elements, such as O, C, Ne, Ca, Mg and Ni, but in very small amounts compared to Type II SNe. We assume the stellar yields for Type Ia SNe from Iwamoto et al. (1999).

Massive stars ($8 \; M_{\odot} \; < \; M \; \leq \; 100 \; M_{\odot}$) are the progenitor of  core-collapse SNe which can be either Type II SNe or Type Ib/c SNe. These latter can arise from binary systems or Wolf-Rayet stars whereas Type II SNe originate from the massive stars in the lower mass range. Type II SNe  mainly produce the so called $\alpha$-elements, such as O, Mg, Ne, Ca, S and Si and Ti, but also some Fe and Fe-peak elements although in smaller amounts than Type Ia SNe.  We adopt here the stellar yields for massive stars by Woosley \& Weaver (1995) with the suggested modifications of Fran\c cois et al. (2004). 
However, the most important modifications concern some Fe-peak elements, except Fe itself, whereas for the $\alpha$-elements, with the exception of Mg which has been increased relative to the original yields, the yields are substantially unmodified.
The modifications of the yields in Fran\c cois et al. (2004) were required to fit at best and at the same time the [element/Fe] versus [Fe/H] patterns and the solar absolute abundances. We keep the same prescriptions here with the purpose of testing the infall laws without changing the other model parameters. 

Finally, we start with primordial gas and the assumed primordial abundances of D and $^{3}$He we have chosen: $3.90 \cdot 10^{-5}$ and $2.25 \cdot 10^{-5}$, respectively. The reference solar abundances are those by Asplund et al. (2005).  

\section{The model by Chiappini et al. (1997)}

Prior to the two-infall model of Chiappini et al. (1997), different models assuming gas accretion onto the galactic disk had been constructed. For example, dynamical models, such as the one of Larson (1976), viscous models (Lacey \& Fall 1985; Sommer-Larsen \& Yoshii 1989, 1990; Tsujimoto et al. 1995), inhomogeneous models (Malinie et al. 1993), detailed chemical evolution models (Matteucci \& Greggio 1986; Tosi 1988; Matteucci \& Fran\c cois 1989; Pagel 1989; Matteucci \& Fran\c cois 1992; Carigi 1994; Giovagnoli \& Tosi 1995; Ferrini et al. 1994; Pardi \& Ferrini 1994; Pardi, Ferrini \& Matteucci 1995; Prantzos \& Aubert 1995; Timmes, Woosley \& Weaver 1995) and chemodynamic models (Samland \& Hensler 1996, Burkert, Truran \& Hensler 1992). The model by Chiappini et al. (1997) was the first in which two main infall episodes for the formation of the Galactic components were suggested. In particular, they assumed that 
the first infall episode was responsible for the formation of the halo and thick-disk stars that originated from a fast dissipative collapse. The second infall episode formed the thin-disk component, with a timescale much longer than that of the thick-disk formation. The authors included in the model also a threshold in the gas density, below which the star formation process stops. The existence of such a threshold value is suggested by observations relative to the star formation in external disk galaxies (Kennicutt 1998, but see Boissier et al. 2006). The physical reason for a threshold in the star formation is related to the gravitational stability, according to which, below a critical density, the gas is stable against density condensations and, consequently, the star formation is suppressed.  
In the two-infall model
the halo- thick disk and the thin disk evolutions occur at different rates, mostly as a result of different accretion rates. 
With these precise prescriptions it is possible to reproduce the majority of the observed properties of the Milky Way and this shows how important is the choice of the accretion law for the gas coupled with the star formation rate in the Galaxy evolution.

In the model by Chiappini et al. (1997) the Galactic disk is approximated by a series of concentric annuli, 2 kpc wide, without exchange of matter between them. The basic
equations are the same as in Matteucci \& Fran\c cois (1989). The two main differences between the model by Chiappini et al. (1997) and Matteucci \& Fran\c cois (1989) are the rate of mass accretion and the rate of star formation. 
Moreover, in the model by Chiappini et al. (1997) the material accreted by the Galactic thin disk comes mainly from extragalactic sources. 
These extragalactic sources could include, for instance, the Magellanic Stream or a major accretion episode (see Beers \& Sommer-Larsen 1995 and references therein).
The two models have in common the ``inside-out'' formation of the thin disk, in the sense that both assume that the timescale for the disk formation increase with galactocentric distance (see section 5). This choice was dictated by the necessity of reproducing the abundance gradients along the Galactic disk.

The SFR is a Schmidt (1955) law with a dependence on the surface gas 
density ($k=1.5$, see Kennicutt 1998) and also on the total surface mass density
(see Dopita \& Ryder 1994).
In particular, the SFR is based on the law originally suggested by Talbot \& Arnett (1975) and then adopted by Chiosi(1980):

\begin{equation}
\psi(r,t)=\nu\left(\frac{\Sigma(r,t) \Sigma_{gas}(r,t)}{\Sigma(r_{\odot},t)^{2}}\right)^{(k-1)}\Sigma_{gas}(r,t)^{k}
\end{equation}

where the constant $\nu$ is a sort of efficiency of the star formation process and is expressed in $Gyr^{-1}$: in particular,  $\nu= 2 \,Gyr^{-1}$ for the halo and $1 \, Gyr^{-1}$ for the disk ($t\ge 1\, Gyr$).
The total surface mass density is represented by $\Sigma(r,t)$, whereas $\Sigma(r_{\odot},t)$ is the total surface mass density at the 
solar position, assumed to be $r_{\odot}=8$ kpc  
(Reid 1993). The quantity $\Sigma_{gas}(r,t)$ represents the 
surface gas density and $t$ represents the time. 
These choices of values for the parameters allow the model to fit very well the observational constraints, in particular in the solar vicinity. A threshold gas density for the star formation in the disk of $7 M_{\odot} pc^{-2}$ is adopted in all the models presented here.

The IMF is that of Scalo (1986) normalized over a mass range of 
0.1-100 $M_{\odot}$and it is assumed to be constant in space and time.

\section{The cosmological simulation}

The main aim of our work is to follow the chemical evolution of spiral
galaxies in a cosmological context. To this aim, we run a dark matter-only
cosmological simulation, using the public tree-code GADGET2 (Springel 2005),
in order to produce and study dark matter halos in which a spiral galaxies
can form. Our simulated box has a side of 24 $h^{-1}$ Mpc. We used $256^{3}$
particles.  We adopted the standard cosmological parameters from WMAP 3-years
(Spergel et al. 2007), namely $\Omega_{0} = 0.275$, $\Omega_{\lambda} = 0.725$
and $\Omega_{b} = 0.041$. Every DM particle has a mass equal to $6.289 \cdot
10^{7} h^{-1} M_{\odot}$ and the Plummer-equivalent softening length is set to
3.75 $h^{-1}$ comoving kpc till redshift $z=2$ and to $1.25 \,h^{-1}$
physical kpc since $z=2$. We use the public package GRAFIC (Bertschinger 1995)
to set up our initial conditions. The simulation started at redshift $z=20$
and 28 outputs were produced. We have chosen to use a quite large spread in
the redshifts at the beginning, while in the last part of the simulation, where
a small change in the redshift corresponds a large change in time, the
redshifts are closer.
We checked that the final mass function of DM halos and the power spectrum
are in agreement with theoretical expectations.

\begin{figure*}
\begin{center}
\includegraphics[width=0.4\textwidth]{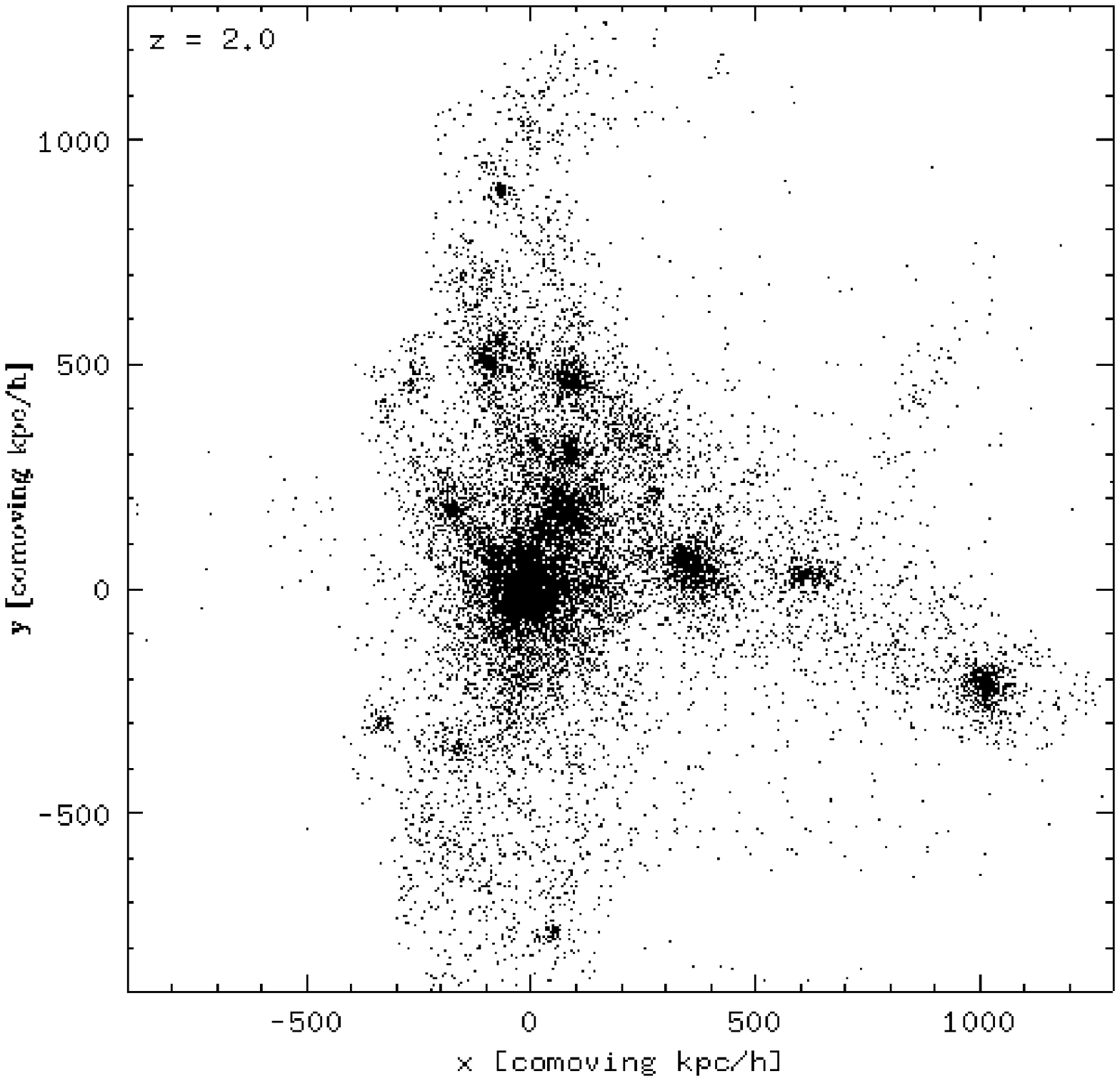}
\includegraphics[width=0.4\textwidth]{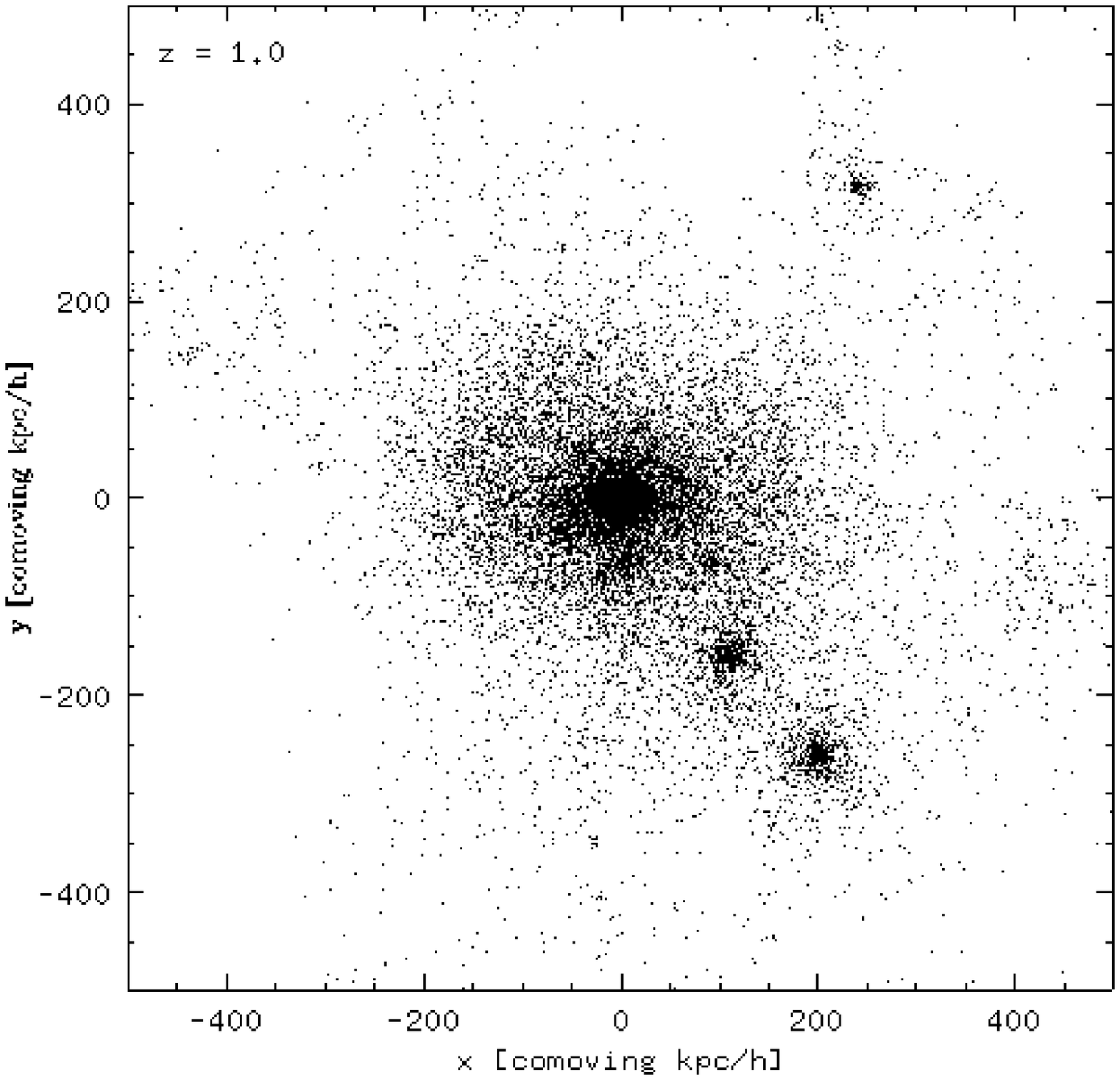}

\includegraphics[width=0.4\textwidth]{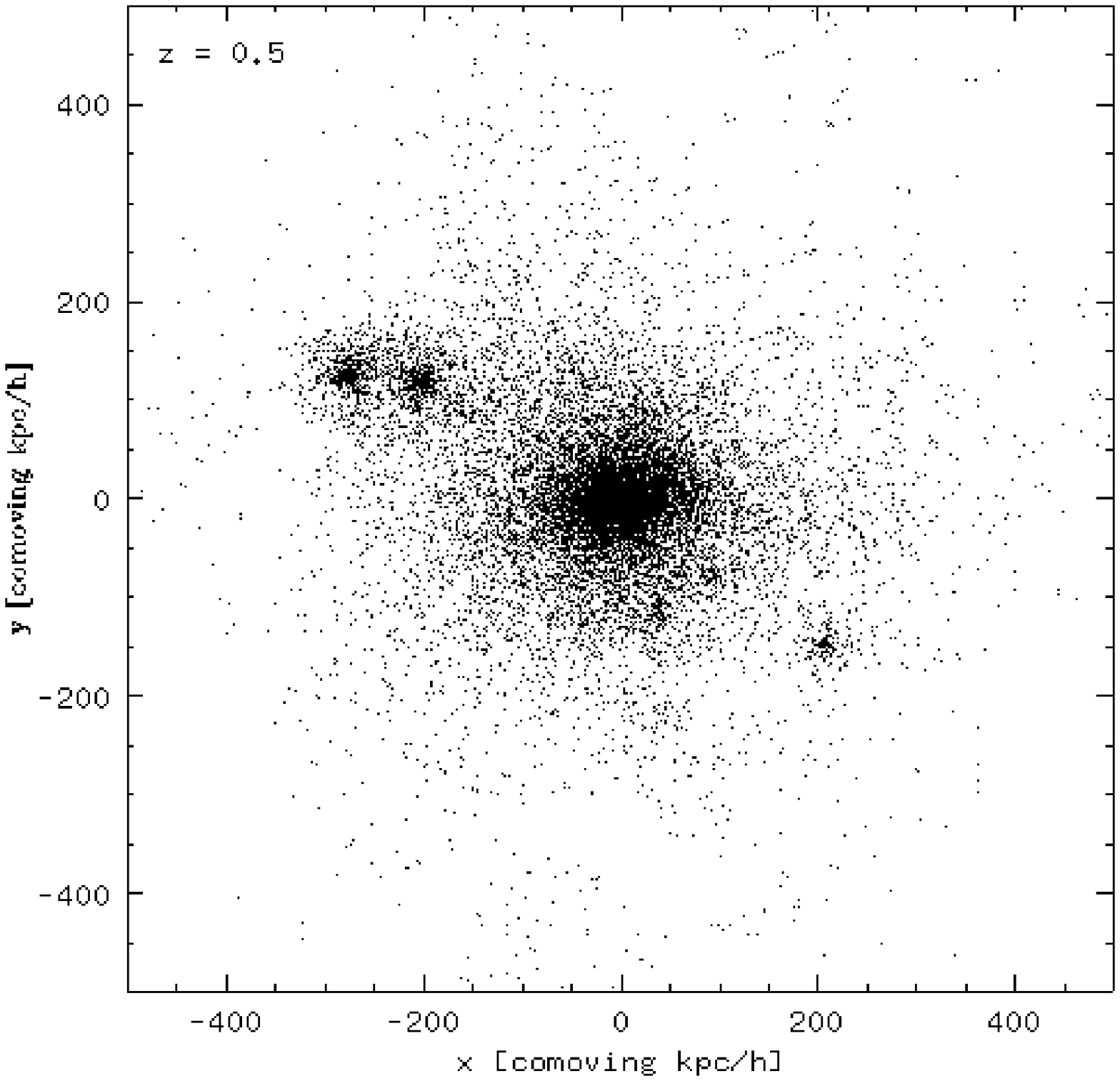}
\includegraphics[width=0.4\textwidth]{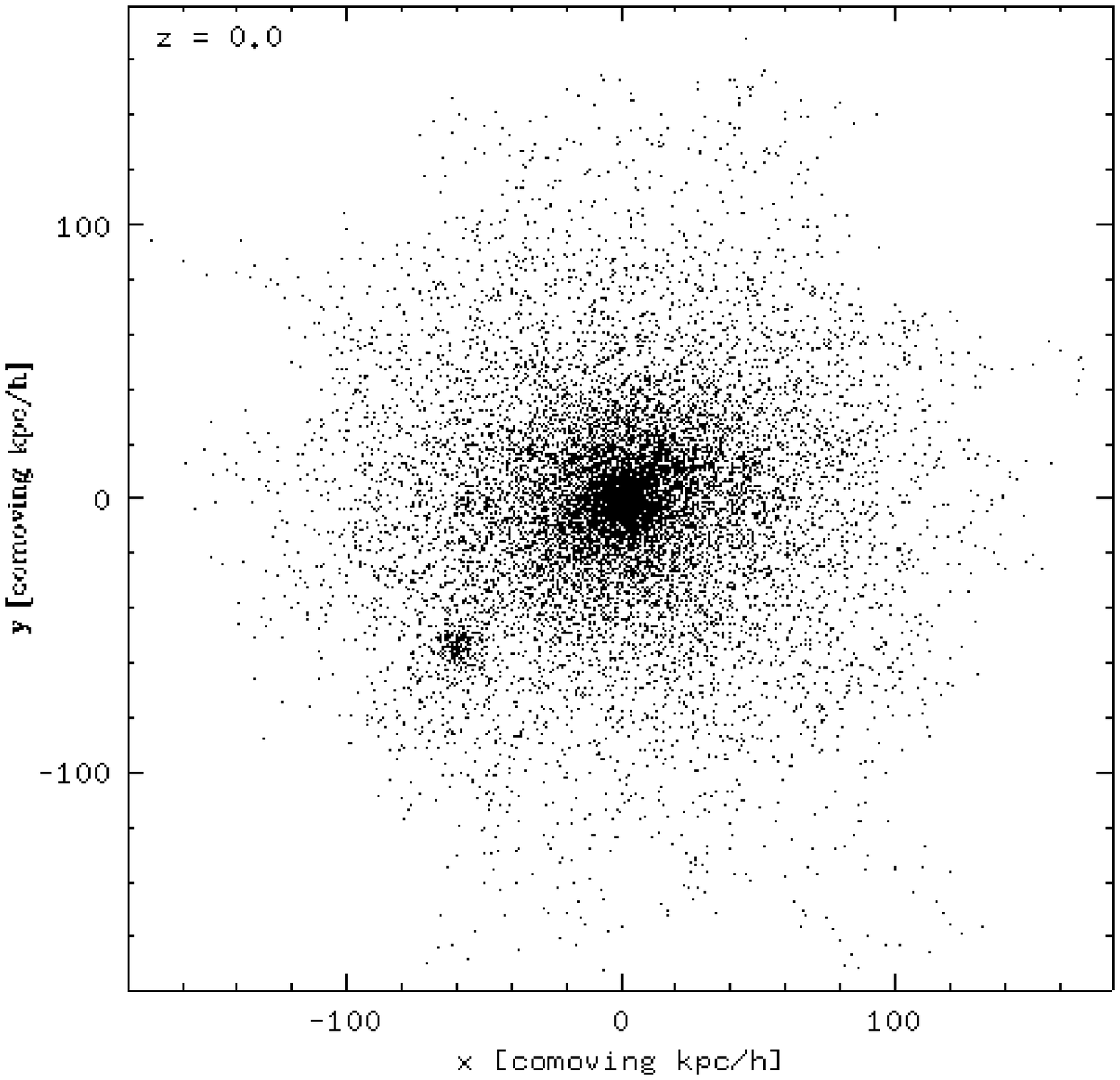}
\caption{This figure represents our best cosmological halo, i.e. halo 48001, at four different redshifts (z = 2.0, z = 1.0, z = 0.5 and z = 0.0).}
\end{center}
\end{figure*}

We identified DM halos at redshift $z=0$ using a standard Friend-of-Friends
algorithm, with a linking length $l=0.17$ mean (comoving) interparticle
distance. After that, we determined the virial mass and radius for each DM 
halo, using the center of mass of the F-o-F group as the halo center. Here we 
define the virial radius as the radius of the sphere within which the matter density contrast is $\delta \approx 100$ times the critical density, with $\delta$
given by the cosmological parameter as in Navarro \& Steinmetz (2000).

We then built the mass accretion history of our halos. To achieve this goal,
we analysed 28 outputs from redshift $z = 9.0$ to $z = 0$. We identified all DM
halos in each snapshot using the procedure sketched above, except for the
fact that we used the redshift-dependent density contrast given by Bryan \&
Norman (1997) to define the virial radius as a function of $z$.  At any output
$z_{i+1}$, we found all the progenitors of our halos at redshift
$z_i$. We defined a halo at redshift $z_{i+1}$ to be a progenitor of one at
$z_i$ if at least 50\% of its particles belong to the candidate offspring (see
e.g. Kauffmann 2001, Springel et al. 2001 for a discussion of this threshold).
The mass accretion history is defined as the mass of the main progenitor of
the halo as a function of redshift.  Having the mass accretion histories,
we were able to identify the redshift of formation (defined as the epoch at
which half of the mass of the forming halos were accreted) and the
redshift at which each halo experienced its last major merger (defined as an
increase of at least 25\% of its mass with respect to the mass of its main
progenitor at the previous redshift).
To identify  the DM halos which can host a spiral galaxy similar to the MW we
used selection criteria based on four different characteristics of the
halos:
\begin{itemize}
 \item mass between $5 \cdot 10^{11} M_{\odot}$ and $5 \cdot 10^{12}
   M_{\odot}$;
\item spin parameter $\lambda > 0.04$; 
\item redshift of last major merger larger than $z=2.5$;
\item redshift of formation larger than $z=1.0$.
\end{itemize}

We found four DM halos compatible with our selection criteria. We label them
with their F-o-F group number, i.e. group 48001, group 52888, group 56004 and
group 6460. We note that, given our simulated volume, the expected number of
halos in our mass range is higher: using a Press \& Schecter mass function,
approximately 70 halos are expected.
However, the requirement of having a ``quiescent'' formation history and
a high spin parameter greatly reduces their number (see e.g. D'Onghia \&
Burkert (2004) and references therein for a discussion on this point). In this
paper we want to focus on the chemical evolution of a MW-like galaxy in its
cosmological context, and therefore we will not discuss issues connected with
the angular momentum problem which arises when performing a direct simulation
of the formation of a disk-like galaxy in a cosmological dark matter halo. 
Also, we could have
obtained a larger number of halos by relaxing the third of the above
constraints, but for the purpose of the present work it is more important to focus
on the most promising DM halos than obtaining statistics.
So, we simply used the (few) best candidates as example halos. 

Assuming that the baryonic matter follows the same accretion pattern 
of the dark matter, and that it represents 
the 19\%
(the cosmological baryon fraction) of all the infalling matter, we obtained a 
final baryonic mass for the Galaxy of $1.7 \cdot 10^{11} M_{\odot}$.
This approach is similar to that followed by Robertson et al. (2005) except 
that we did not make any hypothesis on the fraction of cold gas falling into 
the disk but we used the observations to fix it.
In this way, we obtained the
baryon infall law from the mass accretion history of each halo.

Here, we do not make any attempt to model the disk formation inside the hierarchically growing DM halo. This is undoubtedly an over-simplification of the physics involved. On the other hand, the issue of disk galaxy formation in hierarchical cosmologies is far from being solved. Any attempt to model the formation of the disk should use a number of assumptions which are currently under debate. As an example, the structure of the disk is obviously driven by the gas cooling coupled with its angular momentum content. Semi-analitical galaxy formation models (SAMs) usually assume that DM and gas share the same specific angular momentum. But this point is very controversial (see e.g. D'Onghia \& Burkert 2004, D'Onghia et al. 2006, and references therein). Even direct self consistent numerical simulations  are not currently  able to solve the problem,  which may  (Governato et al 2007) or may not (Abadi et al 2003) be simply due to insufficient numerical resolution and/or an insufficiently detailed treatment of supernovae feedback.
Lacking a widely accepted model for the formation of the disk, we prefer to keep our model as simple as possible and to verify if the cosmological growth of the halo is compatible with the observational constraints obtained using available data on the chemical composition of stars and gas in the Milky Way.

In particular, we assumed that the derived infall law has the same functional form for the whole Milky Way, but that the normalization constant is different for different Galactic regions. In other words, the normalization constants were obtained by reproducing  the present time total surface mass density at any  specific galactocentric distance (see next section), although  here we will focus on the solar neighbourhood, leaving to a forthcoming paper a more detailed study of the whole Galactic disk.
Finally, we also considered an arithmetic mean of the infall laws of all four
halos, in order to have an ``average'' cosmological infall law to study. 
In table 1 we summarize the characteristics of the halos.
Figure 1 represents our best cosmological halo (halo 48001) at four different redshifts (z = 0.0, z = 0.5, z = 1.0 and z = 2.0).

\begin{table*}
\begin{minipage}{60mm}
\caption{Characteristics of the chosen DM halos}
\end{minipage}
\centering
\begin{tabular}{c|c|c|c|c}
\noalign{\smallskip}
\hline
\hline
\noalign{\smallskip}
Group & Mass [$10^{10} M_{\odot}$] & Spin parameter & Redshift major merger & Redshift of formation \\
\noalign{\smallskip}
\hline
\noalign{\smallskip} 
48001 & 90.26 & 0.045 & 5.00 & 1.75 - 1.50 \\ 
52888 & 465.75 &  0.059 & 3.75 & 1.50 - 1.25 \\ 
56009 & 90.73 & 0.049 & 3.25 & 2.00 - 1.75 \\
6460 & 61.94 & 0.041 & 2.50 & 1.25 - 1.00 \\
\noalign{\smallskip}
\hline
\hline
\end{tabular}
\end{table*}

\section{The infall laws}

In testing the accretion laws, we started by adopting the two-infall law 
model, as suggested by Chiappini et al. (1997).  This law presents two distinct
peaks. During the first peak the halo and thick disk formed whereas during the
second peak the thin disk was assembled.  The two accretion events are
considered to be independent from each other and only a very small fraction of
the gas lost from the halo was assumed to have fallen onto the disk.  The infall
law that we indicate as $A(r,t)$ is expressed as:
\begin{eqnarray}
A(r,t)= a(r) e^{-t/ \tau_{H}(r)}+ b(r) e^{-(t-t_{max})/ \tau_{D}(r)}\nonumber
\end{eqnarray}
\begin{eqnarray}
[M_{\odot} \; pc^{-2} \; Gyr^{-1}] 
\end{eqnarray}
where $a(r)$ and $b(r)$ are two parameters fixed by reproducing the total present time surface mass density along the Galactic disk. In particular, in the solar vicinity the total surface mass density $\Sigma_{tot} = 51 \pm 6 M_{\odot} \; pc^{-2}$  (see Boissier \& Prantzos 1999). $t_{max}=1.0$ Gyr
is the time for the maximum infall on the thin  disk, $\tau_{H}= 2.0$ Gyr
is the time scale for the formation of the halo thick-disk and $\tau(r)$
is the timescale for the formation of the thin disk and it is a function of the galactocentric distance (formation inside-out, Matteucci and Fran\c cois 1989;
Chiappini et al. 2001).
In particular, it is assumed  that:
\begin{equation}
\tau_{D}=1.033 r (kpc) - 1.267 \,\, Gyr
\end{equation}
Besides this infall law, we tested other possible laws, such as a time constant
infall rate.  In particular:
\begin{equation}
A(r, t)=3.80 \, [M_{\odot} \; pc^{-2} \; Gyr^{-1}] 
\end{equation}
This law is probably not realistic although  Naab \& Ostriker 2006 concluded that an almost constant infall law over the disk lifetime was to be preferred. Here we adopted it mainly for the purpose of comparison with more realistic laws. We adopted that particular value of the infall rate in order to reproduce  the present  time infall rate (see Table 3), as well as the present time total surface mass density.

The third infall law we tested it is a linear infall law, given by:
\begin{equation}
A(r,t)= 6.57 - 0.4 \cdot t \, [M_{\odot} \; pc^{-2} \; Gyr^{-1}]
\end{equation}
Again, we used this particular expression in order to reproduce the present time $\Sigma_{tot}$ and infall rate.

The fourth  adopted infall law is the same as that of Chiappini et al. (1997) but with pre-enriched infalling gas. The metallicity of the infalling gas which forms the disk was assumed to be 10 times lower than the present time interstellar medium (ISM) metallicity while the infalling gas which forms the halo is still primordial. The assumed chemical composition of the infalling gas does not assume solar abundance ratios but reflects the composition of the halo-thick disk. 

Then, we tested the infall laws derived from the cosmological simulations done with GADGET2 (Springel 2005), as described before.
In particular, to derive the cosmological infall law we proceeded in the following way:

\begin{equation}
A(r,t)= a(r) 0.19 {dM_{DM} \over dt}[M_{\odot} \; pc^{-2} \; Gyr^{-1}] 
\end{equation}
where 0.19 is the cosmological baryonic fraction and $a(r)$ is a normalization constant fixed to reproduce the present time total surface mass density  along the disk, in analogy with eq. (2). For the solar ring $a(r) = {\Sigma (r_{\odot}, t_G) \over M_{Gal}}$, with $M_{Gal} = 0.19 M_{DM}$ being the baryonic mass of our Galaxy and $t_G$ the Galactic lifetime. In figure 2 we show the values of $a(r)$ versus the galactocentric distance.

\begin{figure}
\begin{center}
\includegraphics[width=0.4\textwidth]{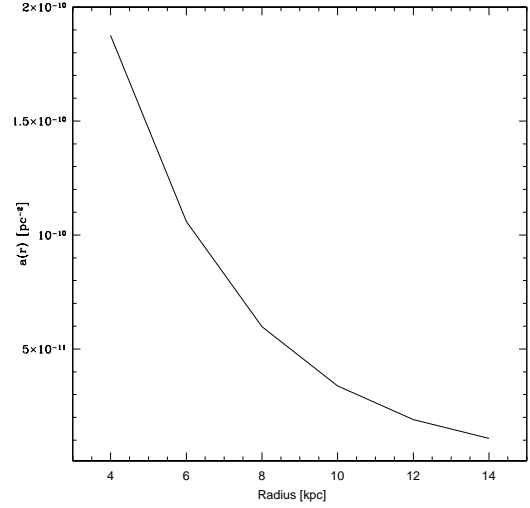}
\caption{$a(r)$ vs radius. This normalization constant is fixed to reproduce the present time total surface mass density along the disk (see eq. 6).}
\end{center}
\end{figure}

One infall law is given by the arithmetic average of the infall laws derived for the four halos and the last infall law is that suggested by Naab \& Ostriker (2006). 
In Table 2 we show the model parameters. The different models are identified mainly by their infall histories.

Our infall laws for the solar region (8 kpc from the Galactic center) are shown in figure 3, whereas in figure 4 we show the
increase in time of the total surface mass density obtained by the mass accretion history of the simulated halos.
It is worth noting that the infall law derived for the best halo
selected as representative of the Milky Way halo is very similar to
the two-infall law by Chiappini et al. (1997). 

We selected our best halo by choosing the one which has a very high redshift of last major merger. This is to ensure the right spin parameter for a Milky Way-like galaxy. The assembly history of this particular halo presents two distinct accretion peaks which produce an infall law very similar to the two-infall model by Chiappini et al. (1997). The only difference with the two-infall model is that in this case the two peaks are placed at a lower redshifts. 
After the two main peaks there are others smaller peaks. The
remarkable fact is that all models predict a present time infall rate
which is in good agreement with the observed one, as quoted by Naab \&
Ostriker (2006). So we can say that according to the infall laws
derived from cosmological simulations the Galaxy had some large infall
episodes at high redshift, followed by smaller ones.

In figure 4 we present the total surface mass density $\Sigma_{tot}$,
expressed as $M_{\odot}pc^{-2}$, as a function of time for all the
models. Once again Models 1 and 4 (two-infall model with primordial
and enriched infall, respectively) have the same $\Sigma_{tot}$. The
linear model predicts the larger final amount of matter, equal to
51.88 $M_{\odot} \; pc^{-2}$. The constant model has a linear growth
(in this case $\Sigma_{tot}$ is the integral of a constant infall law)
and produces 49.98 $M_{\odot} \; pc^{-2}$. Model 10, i.e. the model by
Naab \& Ostriker (2006), is the only one which starts to increase the
amount of matter very slowly (in the solar neighbourhood). After 5 Gyr
from the Big Bang it only has reached 6.00 $M_{\odot} \; pc^{-2}$. The
cosmological models produce results which are quite similar to the
two-infall model. At the beginning their growth is slower but after
$\sim$ 3 - 3.5 Gyr their $\Sigma_{tot}$ increases with a steeper
slope, due to the peaks in the infall law.
   
In figure 5 we show the infall law derived from Model 5 (our best halo) for three galactocentric distances (4, 8 and 14 kpc). As one can see the accretion histories are different at different galactocentric distances, although no assumptions are present about the timescales of disk formation at any radius. This particular behaviour of the infall law with radius needs to be tested and this will be the subject of a forthcoming paper. At the moment we only checked the gradient of O along the galactic disk which is predicted to be very similar to the one obtained with the two-infall model for $R_G> 8$ kpc, whereas it is flatter for $R_G \le 8$ kpc (see Section 6). Clearly the formation of the bulge is included in the accretion history of the first 2 kpc.

\begin{table*}
\begin{minipage}{120mm}
\caption{Model parameters. In the first column
there is the number of the model, in the second one the adopted infall
law, in the third the time scale for the halo, in the fourth that for
the disk and in the fifth the type of infalling gas. All the models adopt a threshold gas density for star formation in the disk of $7 M_{\odot} pc^{-2}$. The model by Naab \& Ostriker (2006) is the only one which has the threshold also during the formation of the halo. Note that our best cosmological model is Model 5.}
\end{minipage}
\centering
\begin{tabular}{c|c|c|c|c}
\noalign{\smallskip}
\hline
\hline
\noalign{\smallskip}
Model & Infall law & $\tau$ halo & $\tau$ disk & Gas t = 0 \\
 & $[M_{\odot}pc^{-2}Gyr^{-1}]$ & [Gyr] & [Gyr] & \\
\noalign{\smallskip}
\hline
\noalign{\smallskip} 
1 & Two-infall law & 0.8 & 7 & Primordial \\ 
2 & $3.80$ & 0.8 & 7 & Primordial \\ 
3 & $6.57-0.40 \cdot T$ & 0.8 & 7 & Primordial \\
4 & Two-infall law & 0.8 & 7 & Enriched (1/10 $Z_{today}$) \\
5 & Group 48001 & - & - & Primordial \\
6 & Group 52888 & - & - & Primordial \\
7 & Group 56009 & - & - & Primordial \\
8 & Group 6460 & - & - & Primordial \\
9 & Mean & - & - & Primordial \\
10 & Naab \& Ostriker  & - & - & Primordial \\
\noalign{\smallskip}
\hline
\hline
\end{tabular}
\end{table*}

\section{Results}
In this section we present the chemical evolution results.

Some results are shown in Tables 3, 4 and 5. In particular, in Table 3
we show the predicted present star formation rates, the present
infall and the present SNIa and SNII rates, compared with the
corresponding observational values.  In Table 4 we plot the total
amount of gas and stars, the $\Sigma_{gas}/\Sigma_{tot}$ and the total
surface mass density in a ring of 2 kpc centered at the Sun's
galactocentric distance (8 kpc).  Finally, Table 5 presents the
predicted solar absolute abundances by mass for Fe, C, Mg, N, O and
Si, namely the abundances in the ISM at the time of birth of the solar
system 4.5 Gyr ago, compared with the observed ones by Asplund et
al. (2005).

\begin{table*}
\begin{minipage}{120mm}
\caption{Present time values for all the models and observed values as reported in Boissier \& Prantzos (1999) and Chiappini et al. (2001).}
\end{minipage}
\centering
\begin{tabular}{c|c|c|c|c}
\noalign{\smallskip}
\hline
\hline
\noalign{\smallskip}
Model & SFR & Infall & SNII rate & SNIa rate\\
 & $[M_{\odot}pc^{-2}Gyr^{-1}]$ & $[M_{\odot}pc^{-2}Gyr^{-1}]$ & $[pc^{-2}Gyr^{-1}]$ & $[pc^{-2}Gyr^{-1}]$ \\
\noalign{\smallskip}
\hline
\noalign{\smallskip}
1 & 2.66 & 1.100 & 0.00900 & 0.00330 \\
2 & 4.55 & 3.800 & 0.01928 & 0.00411 \\
3 & 2.81 & 1.320 & 0.01194 & 0.00391 \\
4 & 2.66 & 1.100 & 0.00900 & 0.00332 \\
5 & 2.65 & 0.528 & 0.00584 & 0.00366 \\
6 & 2.69 & 2.273 & 0.01140 & 0.00347 \\
7 & 2.65 & 0.126 & 0.00229 & 0.00366 \\
8 & 4.01 & 0.998 & 0.01712 & 0.00412 \\
9 & 2.69 & 0.979 & 0.01147 & 0.00381 \\
10 & 4.72 & 3.406 & 0.02002 & 0.00397 \\
\noalign{\smallskip}
\hline
\noalign{\smallskip}
Boissier \& Prantzos (1999) & 2-5 & 1.0-3.3 & 0.02 & 0.0042 $\pm$ 0.0016 \\
\noalign{\smallskip}
\noalign{\smallskip}
\hline
\noalign{\smallskip}
Chiappini et al. (2001) & 2.6 & 1.0 & 0.008 & 0.004 \\
\noalign{\smallskip}
\hline
\hline
\end{tabular}
\end{table*}

\begin{table*}
\begin{minipage}{120mm}
\caption{Present time values for all the models and observed values as reported in Boissier \& Prantzos (1999) and Chiappini et al. (2001)}
\end{minipage}
\centering
\begin{tabular}{c|c|c|c|c}
\noalign{\smallskip}
\hline
\hline
\noalign{\smallskip}
Model & Gas & Stars & $\frac{\Sigma_{gas}}{\Sigma_{tot}}$ & Total \\
 & [$M_{\odot}pc^{-2}$] & [$M_{\odot}pc^{-2}$] & & [$M_{\odot}pc^{-2}$] \\
\noalign{\smallskip}
\hline
\noalign{\smallskip}
1 & 7.00 & 35.24 & 0.1444 & 48.46 \\
2 & 10.11 & 35.09 & 0.2024 & 49.98 \\
3 & 7.42 & 38.66 & 0.1431 & 51.88 \\
4 & 7.00 & 35.24 & 0.1444 & 48.46 \\
5 & 6.99 & 36.13 & 0.1439 & 48.55 \\
6 & 7.06 & 35.60 & 0.1455 & 48.53 \\
7 & 7.00 & 36.69 & 0.1442 & 48.56 \\
8 & 9.21 & 34.75 & 0.2056 & 48.55 \\
9 & 7.06 & 36.32 & 0.1455 & 48.55 \\
10 & 10.29 & 34.29 & 0.2099 & 49.04 \\
\noalign{\smallskip}
\hline
\noalign{\smallskip}
Boissier \& Prantzos (1999) & 13 $\pm$ 3 & 35 $\pm$ 5 & 0.15 - 0.25  & 51 $\pm$ 6 \\
\noalign{\smallskip}
\noalign{\smallskip}
\hline
\noalign{\smallskip}
Chiappini et al. (2001) & 7.0 & 36.3 & 0.13 & 53.85 \\
\noalign{\smallskip}
\hline
\hline
\end{tabular}
\end{table*}

\begin{table*}
\begin{minipage}{120mm}
\caption{Predicted and observed solar abundances by mass (after 8.64 Gyr from the Big Bang)}
\end{minipage}
\centering
\begin{tabular}{c|c|c|c|c|c|c}
\noalign{\smallskip}
\hline
\hline
\noalign{\smallskip}
Model & Fe & C & Mg & N & O & Si \\
\noalign{\smallskip}
\hline
\noalign{\smallskip}
1 & 0.162E-02 & 0.156E-02 & 0.774E-03 & 0.121E-02 & 0.592E-02 & 0.980E-03 \\
2 & 0.987E-03 & 0.119E-02 & 0.585E-03 & 0.932E-03 & 0.461E-02 & 0.665E-03 \\
3 & 0.149E-02 & 0.135E-02 & 0.778E-03 & 0.105E-02 & 0.602E-02 & 0.940E-03 \\
4 & 0.111E-02 & 0.157E-02 & 0.691E-03 & 0.121E-02 & 0.547E-02 & 0.771E-03 \\
5 & 0.169E-02 & 0.199E-02 & 0.797E-03 & 0.142E-02 & 0.608E-02 & 0.102E-02 \\
6 & 0.917E-03 & 0.140E-02 & 0.604E-03 & 0.105E-02 & 0.483E-02 & 0.653E-03 \\
7 & 0.107E-02 & 0.168E-02 & 0.701E-03 & 0.124E-02 & 0.559E-02 & 0.761E-03 \\
8 & 0.126E-02 & 0.212E-02 & 0.796E-03 & 0.144E-02 & 0.635E-02 & 0.879E-03 \\
9 & 0.111E-02 & 0.173E-02 & 0.716E-03 & 0.127E-02 & 0.570E-02 & 0.783E-03 \\
10 & 0.531E-03 & 0.102E-02 & 0.439E-03 & 0.784E-03 & 0.362E-02 & 0.432E-03 \\
\noalign{\smallskip}
\hline
\noalign{\smallskip}
Asplund \& al. (2005) & 0.116E-02 & 0.217E-02 & 0.601E-03 & 0.623E-03 & 0.540E-02 & 0.669E-03 \\
\noalign{\smallskip}
\hline
\hline
\end{tabular}
\end{table*} 

\begin{figure*}
\centering
\includegraphics[width=0.8\textwidth]{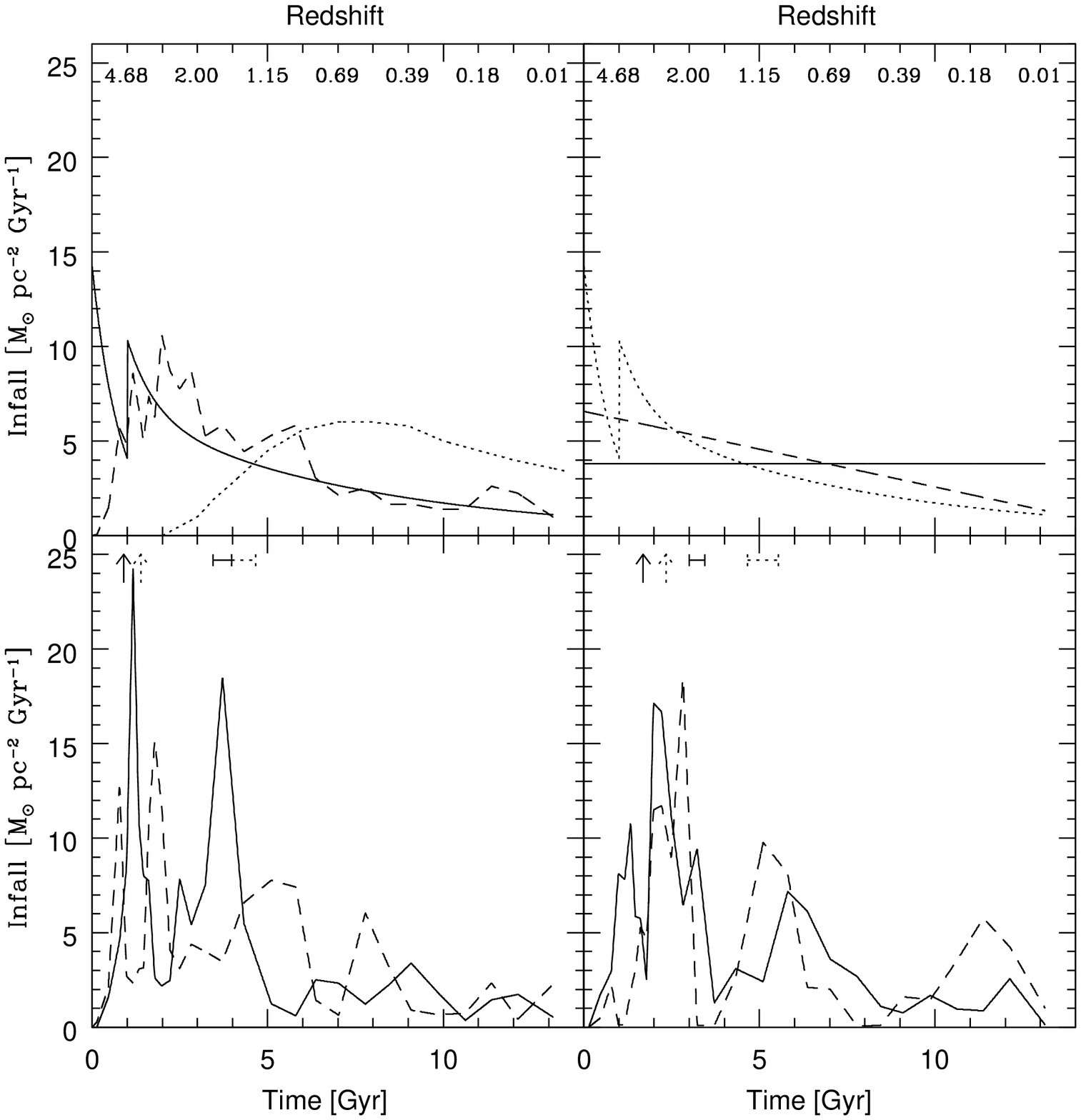}
\caption{Infall vs time. Upper left panel: red solid line is the two-infall model (Model 1); black dashed line is the cosmological mean model (Model 9); green dotted line is the model by Naab \& Ostriker (2006) (Model 10). Upper right panel: magenta solid line is the constant infall model (Model 2); blue dashed line is the linear infall model (Model 3); cyan dotted line is the pre-enriched model ($Z_{inf} = 1/10 \; Z_{today}$, Model 4). Bottom left panel: black solid line is Model 5; magenta dashed line is Model 6. Bottom right panel: blue solid line is Model 7; cyan dashed line is Model 8. In the bottom left panel the black solid arrow represents the redshift of last major merger for Model 5, the magenta dotted arrow the redshift of last major merger for Model 6, the black solid interval the redshift of formation for Model 5 and the magenta dotted interval the redshift of formation for Model 6. In the bottom right panel the blue solid arrow represents the redshift of last major merger for Model 7, the cyan dotted arrow the redshift of last major merger for Model 8, the blue solid interval the redshift of formation for Model 7 and the cyan dotted interval the redshift of formation for Model 8.}
\end{figure*}

\begin{figure*}
\centering
\includegraphics[width=0.8\textwidth]{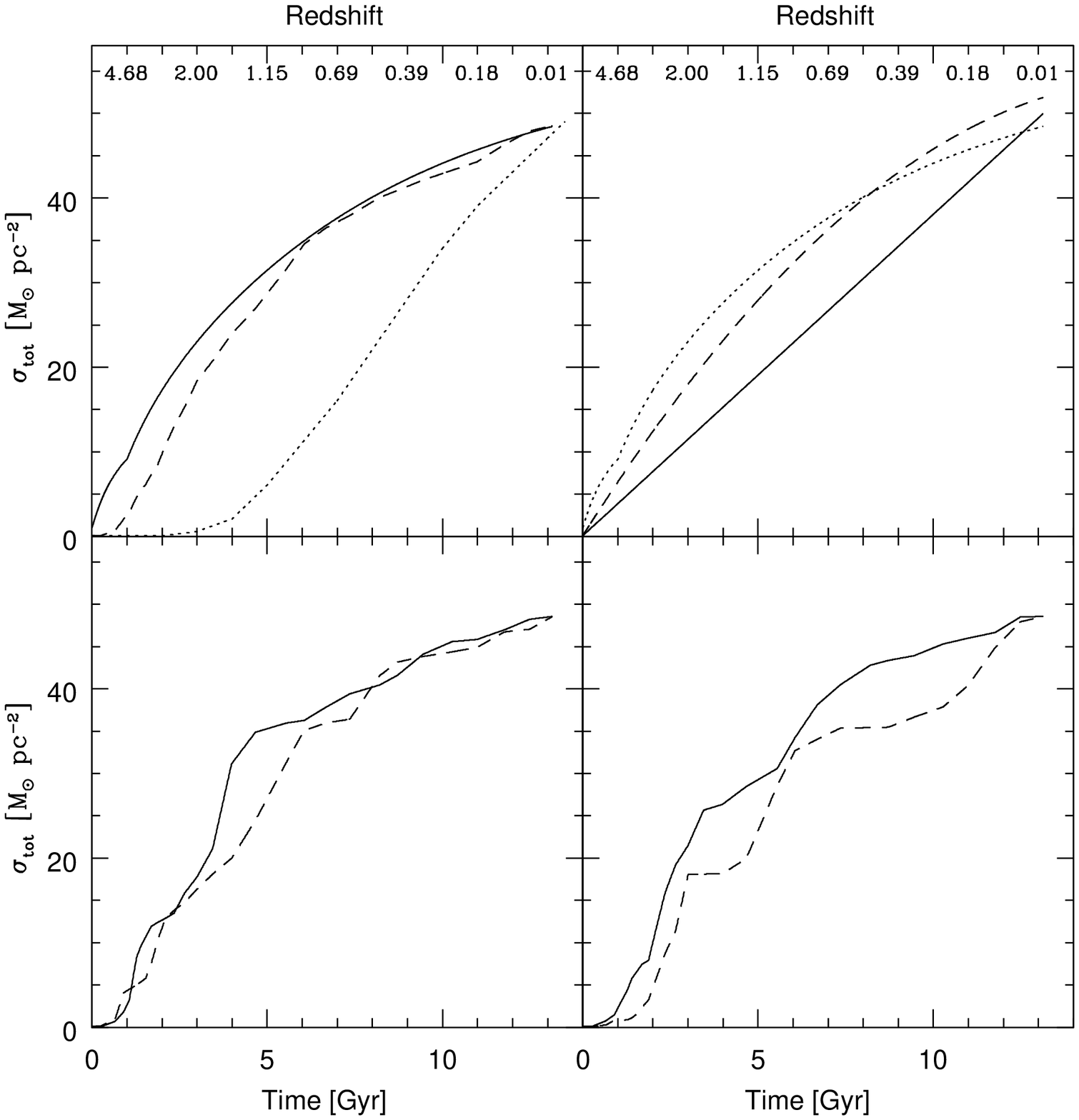}
\caption{$\Sigma_{tot}$ vs time. Upper left panel: red solid line is the two-infall model (Model 1); black dashed line is the cosmological mean model (Model 9); green dotted line is the model by Naab \& Ostriker (2006) (Model 10). Upper right panel: magenta solid line is the constant infall model (Model 2); blue dashed line is the linear infall model (Model 3); cyan dotted line is the pre-enriched model ($Z_{inf} = 1/10 \; Z_{today}$, Model 4). Bottom left panel: black solid line is Model 5; magenta dashed line is Model 6. Bottom right panel: blue solid line is Model 7; cyan dashed line is Model 8.}
\end{figure*}

\begin{figure*}
\centering
\includegraphics[width=0.8\textwidth]{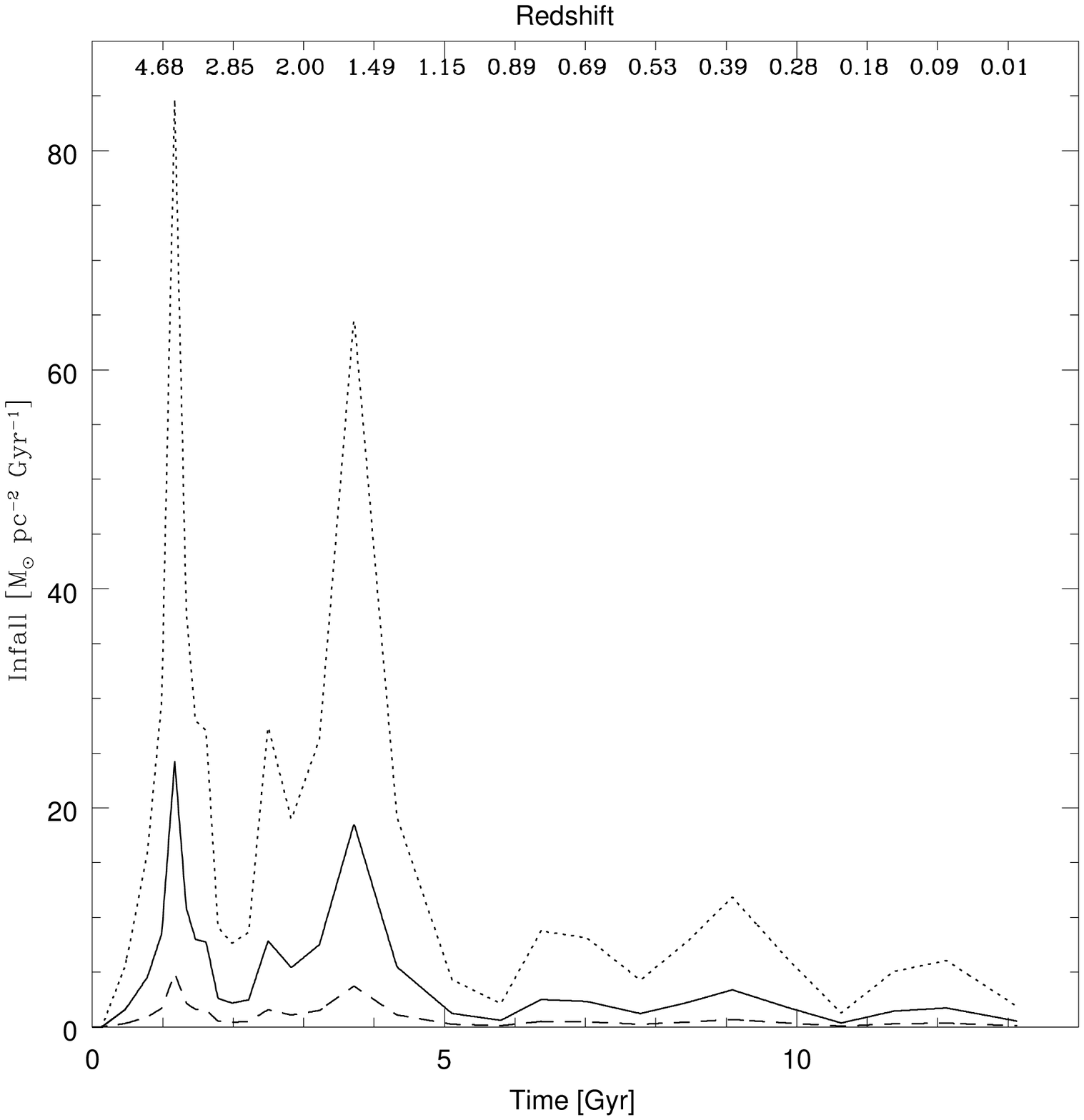}
\caption{This figure represents the infall law our best cosmological halo, i.e. halo 48001,
  at four different radius (4 kpc: blue dotted line; 8 kpc: red solid line; 14 kpc: green dashed line).}
\end{figure*}

Figure 6 shows the star formation rate as a function of cosmic time
for all the models. At high redshift there is a
gap in the SFR for some of the models. This gap is due to the adoption
of a threshold in the surface gas density, below which star formation
does not occur. In all models we have adopted a threshold which is
equal to 7.0 $M_{\odot} \; pc^{-2}$ during the formation of the thin
disk.  Model 10 instead, adopting the infall law suggested by Naab \&
Ostriker (2006), has a star formation threshold equal to 7.0
$M_{\odot} \; pc^{-2}$ both for the halo and the disk.  

From figure 6 we deduce that the constant infall model predicts a growing star
formation rate at low redshifts, a trend which is not predicted by the
other laws. On the other hand, the cosmological best model (Model 5)
predicts a very important peak between 3 and 6 Gyr, which should
correspond to the formation of the bulk of stars in the thin
disk. This peak is directly connected to the trend of the infall
law. After 10 Gyr from the Big Bang the threshold is easily reached in
most of the models, thus causing the SFR to have an oscillating
behaviour.

\begin{figure*}
\centering
\includegraphics[width=0.8\textwidth]{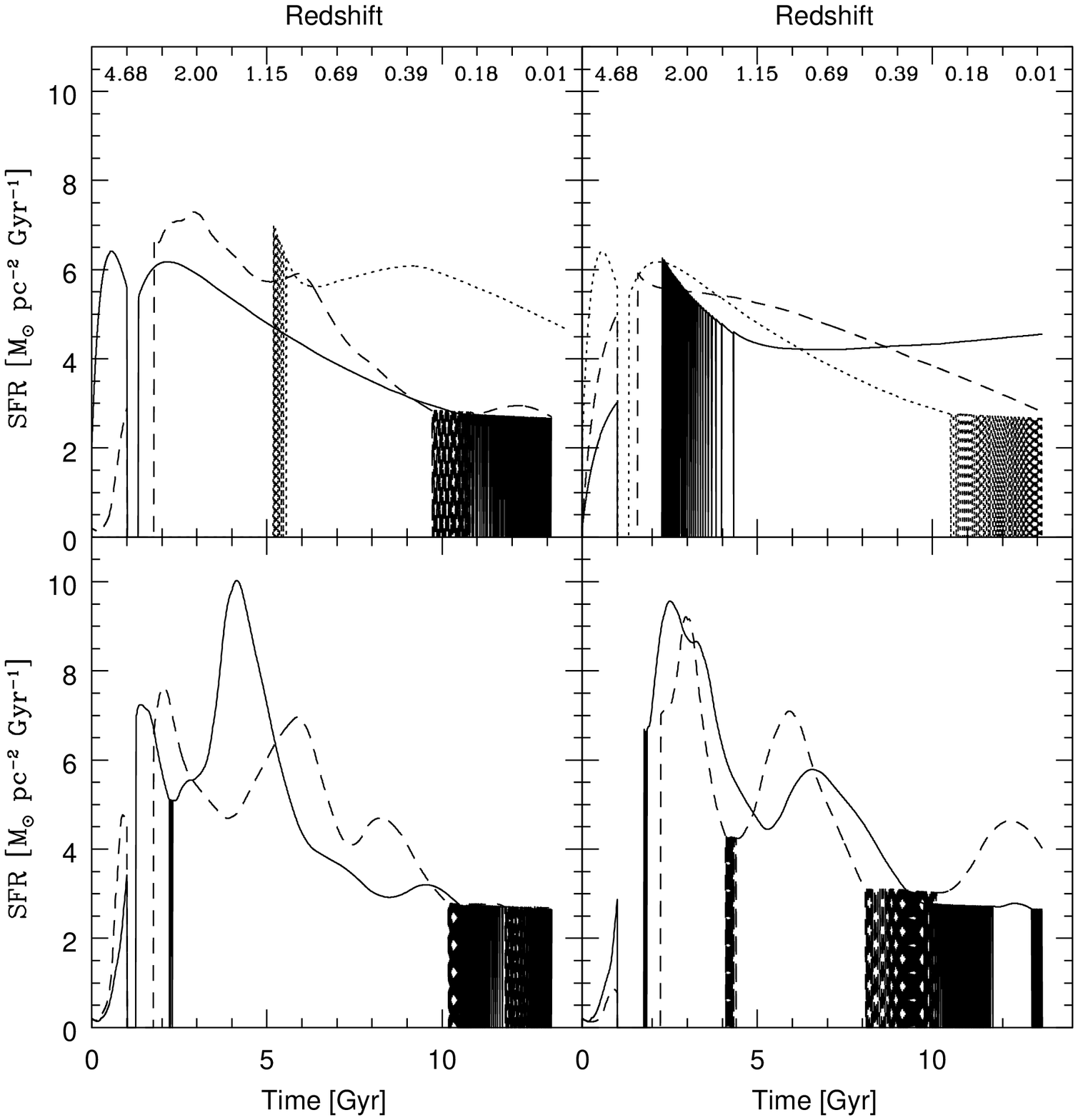}
\caption{SFR vs time. Upper left panel: red solid line is the
two-infall model (Model 1); black dashed line is the cosmological mean
model (Model 9); green dotted line is the model by Naab \& Ostriker
(2006) (Model 10). Upper right panel: magenta solid line is the
constant infall model (Model 2); blue dashed line is the linear infall
model (Model 3); cyan dotted line is the pre-enriched model ($Z_{inf}
= 1/10 \; Z_{today}$, Model 4). Bottom left panel: black solid line is
Model 5; magenta dashed line is Model 6. Bottom right panel: blue
solid line is Model 7; cyan dashed line is Model 8.}
\end{figure*}

In figure 7 we present the SNIa rates for all the models. The
cosmological law of Model 5 predicts a peak for the SNIa rate at about
6 Gyr. This is due to the fact that the SFR in this model has an
important peak at about 5 Gyr. Thanks to this peak, many stars form and
many SNIa explode after a delay of about 1 Gyr. All
the models predict a SNIa rate between 0.003 and 0.004 SNe $pc^{-2} \;
Gyr^{-1}$, in good agreement with the value given by Boissier \&
Prantzos (1999), i.e. 0.0042 $\pm$ 0.0016 .

\begin{figure*}
\centering
\includegraphics[width=0.8\textwidth]{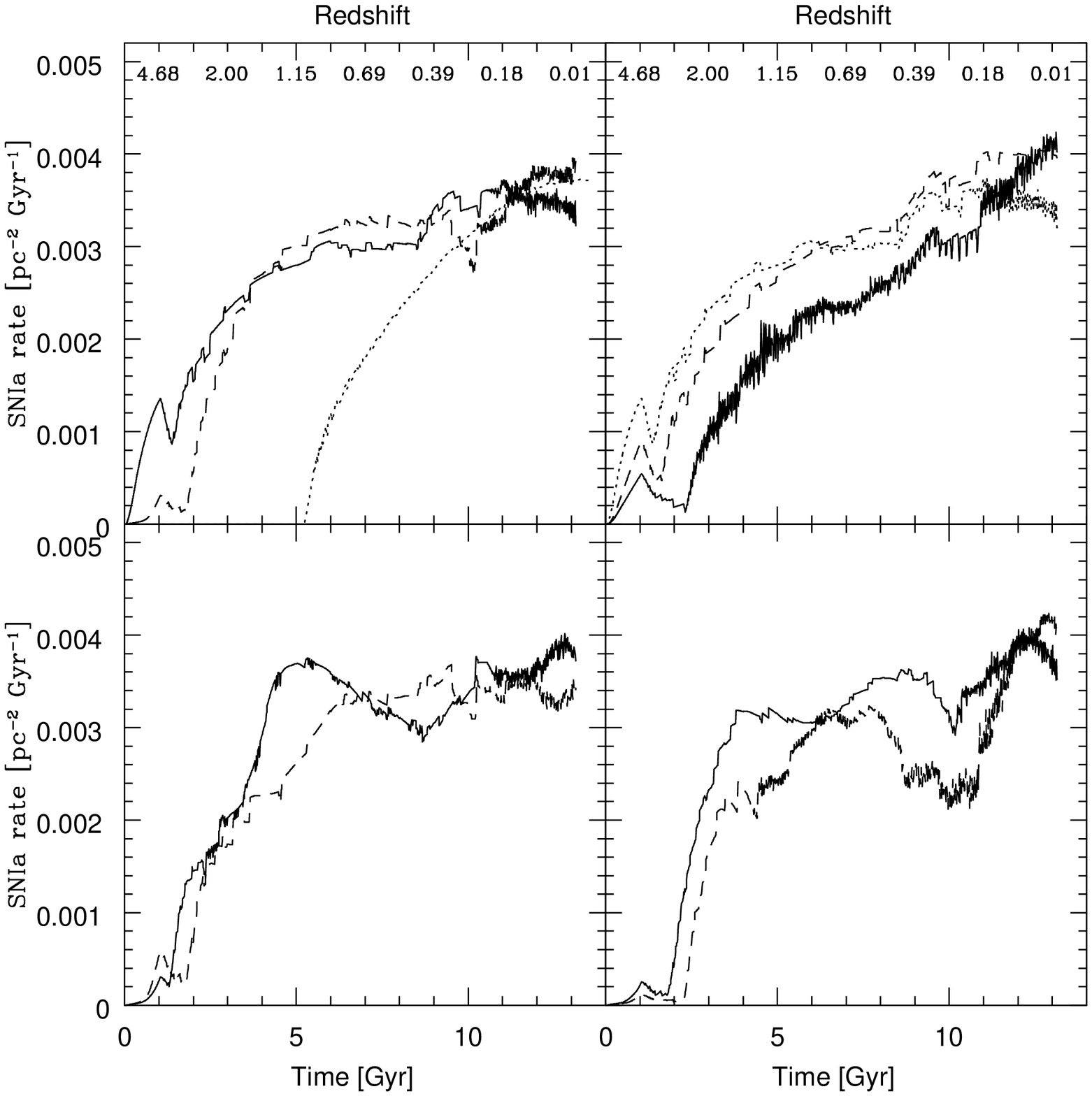}
\caption{SNIa rate vs time. Upper left panel: red solid line is the two-infall model (Model 1); black dashed line is the cosmological mean model (Model 9); green dotted line is the model by Naab \& Ostriker (2006) (Model 10). Upper right panel: magenta solid line is the constant infall model (Model 2); blue dashed line is the linear infall model (Model 3); cyan dotted line is the pre-enriched model ($Z_{inf} = 1/10 \; Z_{today}$, Model 4). Bottom left panel: black solid line is Model 5; magenta dashed line is Model 6. Bottom right panel: blue solid line is Model 7; cyan dashed line is Model 8.}
\end{figure*}

We do not show the rates of SNe II since theirs behaviour is like that of the SFR.
In fact,  Type II SNe are produced by massive stars which live only few millions years. For this reason, the behaviour of the SNII rate is equivalent to that of the SFR. 

In figure 8 we present the predicted [Fe/H] as a function of time for
all models. It is important to note that the model with a constant
infall law (Model 2) and Model 10 never reach the solar abundance.
The reason is that in both models the infall rate during the whole
galactic lifetime is probably overestimated.  In the model by
Chiappini et al. (1997) (our Model 1) [Fe/H] reaches a local peak at
1 Gyr, then decreases slightly to increase again. The little depression
in [Fe/H] is due to the predicted gap in the SFR just before the
formation of the thin disk. In fact, the second infall episode coupled
with the halt in the SF produces a decrease of [Fe/H]. We
can see the same behaviour in the cosmological models.  In particular
in Model 5 the peak is followed by a deeper depression of [Fe/H] and
this is due to the longer gap in the SFR predicted by this model (1-2
Gyr) as opposed to that predicted by Model 1 which is less than 1 Gyr.
This is an important prediction and it can be tested via chemical
abundances. In fact, both Gratton et al. (1996) and Furhmann (1998)
detected such an effect in the [Fe/O] vs. [O/H] and [Fe/Mg]
vs. [Mg/H], respectively.

\begin{figure*}
\centering
\includegraphics[width=0.8\textwidth]{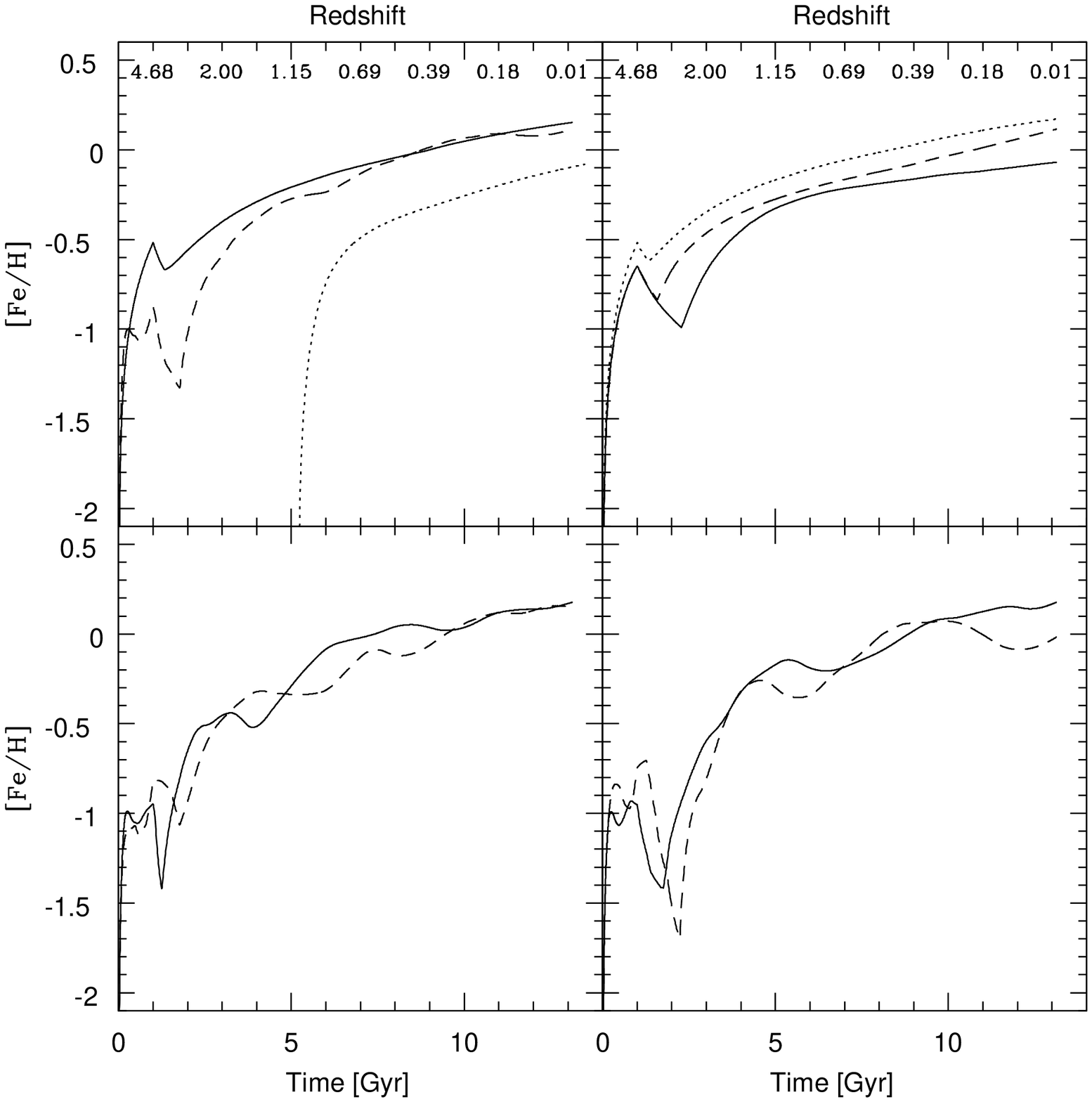}
\caption{[Fe/H] vs time. Upper left panel: red solid line is the two-infall model (Model 1); black dashed line is the cosmological mean model (Model 9); green dotted line is the model by Naab \& Ostriker (2006) (Model 10). Upper right panel: magenta solid line is the constant infall model (Model 2); blue dashed line is the linear infall model (Model 3); cyan dotted line is the pre-enriched model ($Z_{inf} = 1/10 \; Z_{today}$, Model 4). Bottom left panel: black solid line is Model 5; magenta dashed line is Model 6. Bottom right panel: blue solid line is Model 7; cyan dashed line is Model 8.}
\end{figure*} 

A very important constraint for the chemical evolution of the galaxies
is represented by the G-dwarf metallicity distribution. This is the
relative number of G-dwarf stars as a function of [Fe/H]. We have used
the data from Rocha-Pinto \& Maciel (1996), Kotoneva (2002), Jorgensen
(2000) and Wyse (1995). Our predicted metallicity distributions are
shown in figure 9. From this figure, it is clear that Model 10
predicts insufficient high metallicity stars.  On the other
hand, some of the cosmological models such as Model 7 and Model 8
predict too many metal-poor stars. Our best cosmological model,
i.e. Model 5, shows a bimodal metallicity distribution, which is clearly at odds with the data. 

\begin{figure*}
\centering
\includegraphics[width=0.8\textwidth]{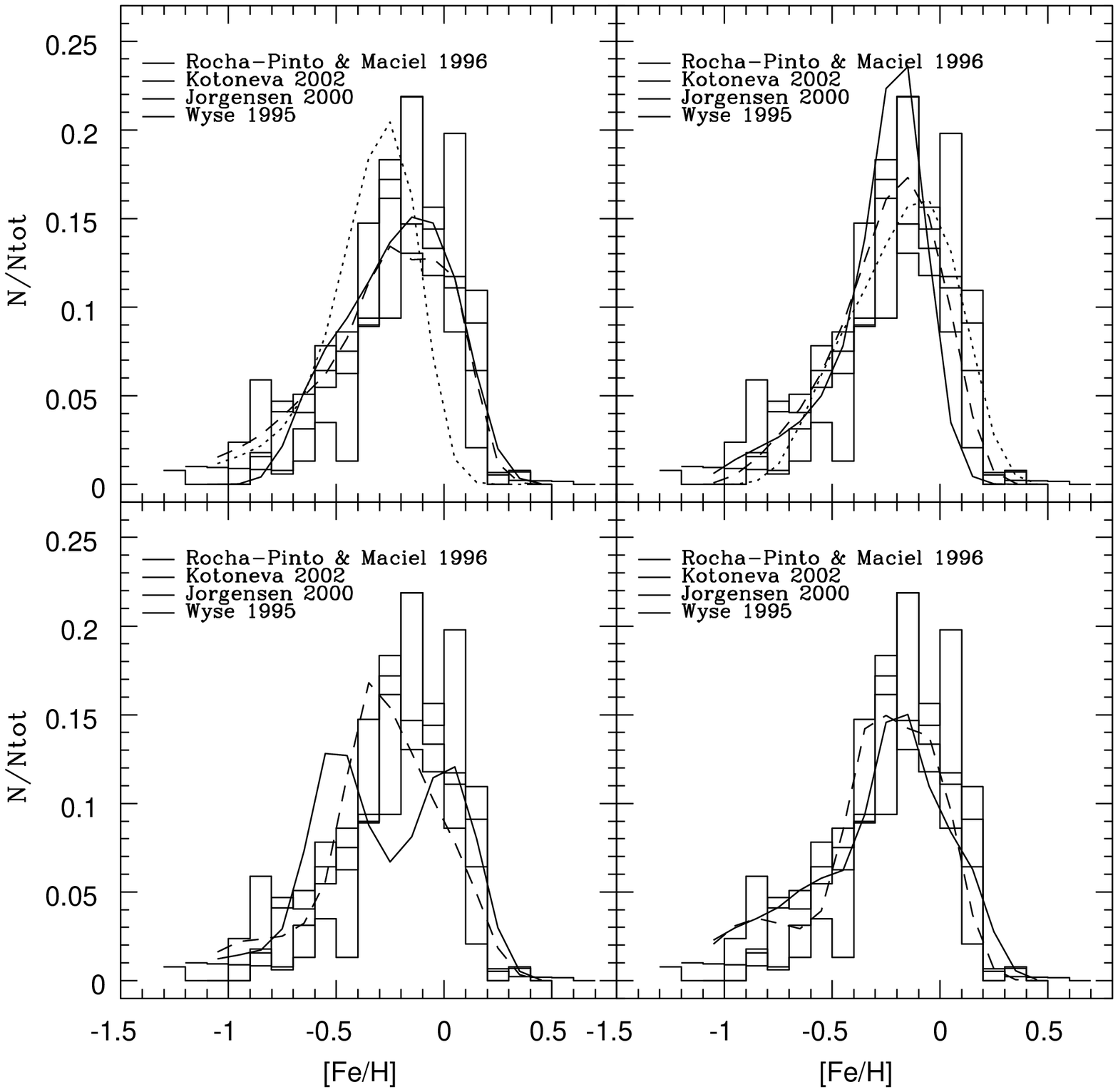}
\caption{G-dwarf metallicity distribution. Upper left panel: red solid line is the two-infall model (Model 1); black dashed line is the cosmological mean model (Model 9); green dotted line is the model by Naab \& Ostriker (2006) (Model 10). Upper right panel: magenta solid line is the constant infall model (Model 2); blue dashed line is the linear infall model (Model 3); cyan dotted line is the pre-enriched model ($Z_{inf} = 1/10 \; Z_{today}$, Model 4). Bottom left panel: black solid line is Model 5; magenta dashed line is Model 6. Bottom right panel: blue solid line is Model 7; cyan dashed line is Model 8.}
\end{figure*} 

The last constraint we study concerns the chemical abundances of several elements, such as O, Mg, Si, N and C. In figure 10 the [O/Fe] as a function of 
[Fe/H] can be seen. Here, the range of [Fe/H] has been restricted to $-2.0$ to $+0.3$ dex in order to see better the predictions relative to the transition between the halo--thick disk and the thin disk. In figure 11 we show the same plots but for the whole range of [Fe/H] down to $-4.0$ dex.

In figure 10 one can see that
cosmological models  have a similar behaviour to the model by Chiappini et al. (1997), except for a longer gap in the SF, which produces a loop in the predicted curves. Such loops arise when SF stops, the $\alpha$-elements are no longer produced whereas Fe continues to be produced. This induces the [O/Fe] to decrease and also the [Fe/H] ratio to decrease to a lesser extent, because of the accretion of primordial gas. Then when SF starts again the [O/Fe] increases again. This loop is very prominant in some models and not in agreement with the data, although some spread is present. 
It is interesting to note that
Model 4, which is the same as  Chiappini et al's model but with the pre-enriched gas, is acceptable.
This is due to the fact that the metallicity of the pre-enriched infalling gas is not so different from the metallicity of the primordial infalling gas.

\begin{figure*}
\centering
\includegraphics[width=0.8\textwidth]{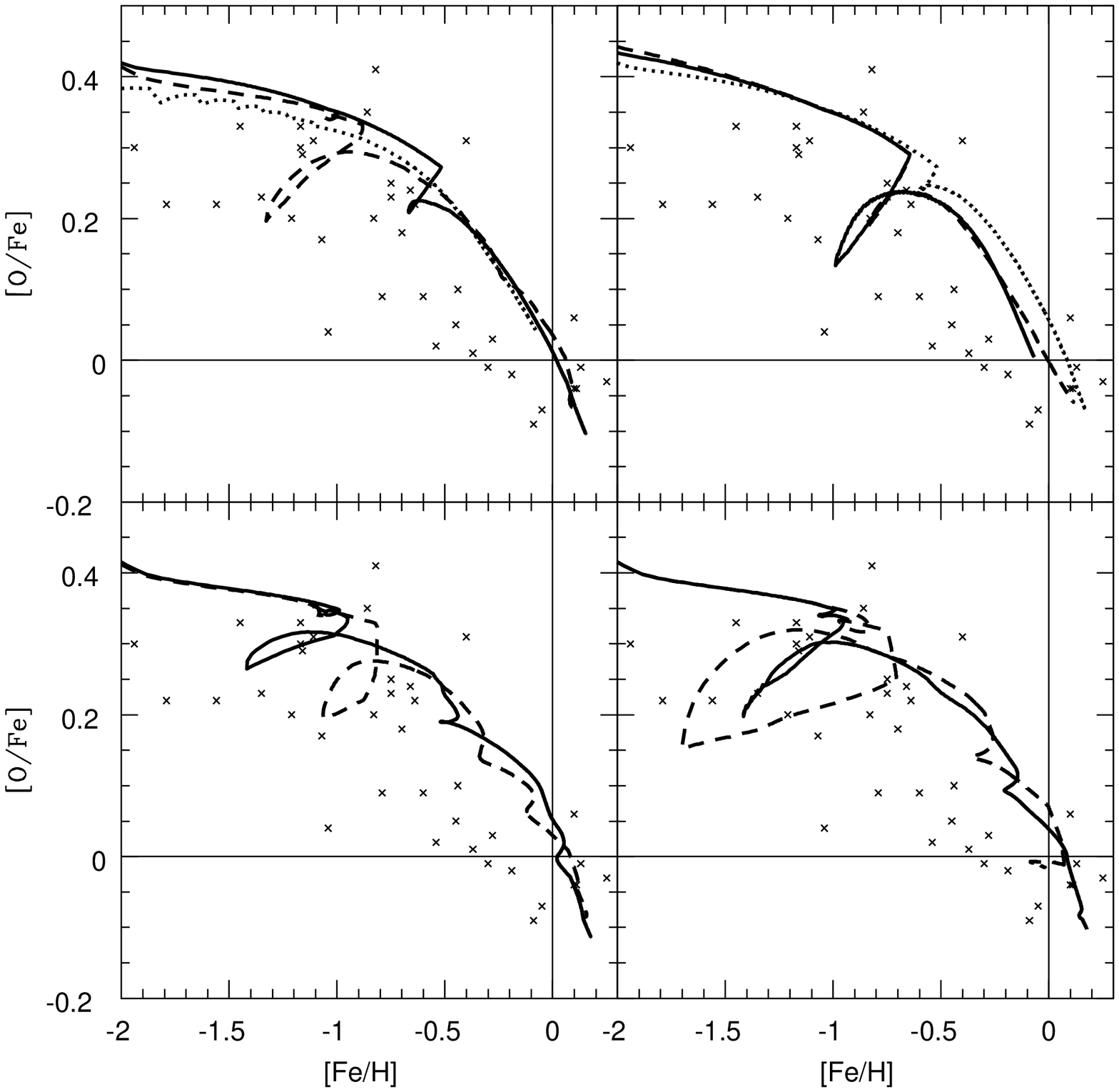}
\caption{[O/Fe] vs [Fe/H]. Upper left panel: red solid line is the two-infall model (Model 1); black dashed line is the cosmological mean model (Model 9); green dotted line is the model by Naab \& Ostriker (2006) (Model 10). Upper right panel: magenta solid line is the constant infall model (Model 2); blue dashed line is the linear infall model (Model 3); cyan dotted line is the pre-enriched model ($Z_{inf} = 1/10 \; Z_{today}$, Model 4). Bottom left panel: black solid line is Model 5; magenta dashed line is Model 6. Bottom right panel: blue solid line is Model 7; cyan dashed line is Model 8. The data are from Fran\c cois et al. (2004) (green crosses).}
\end{figure*}

\begin{figure*}
\centering
\includegraphics[width=0.8\textwidth]{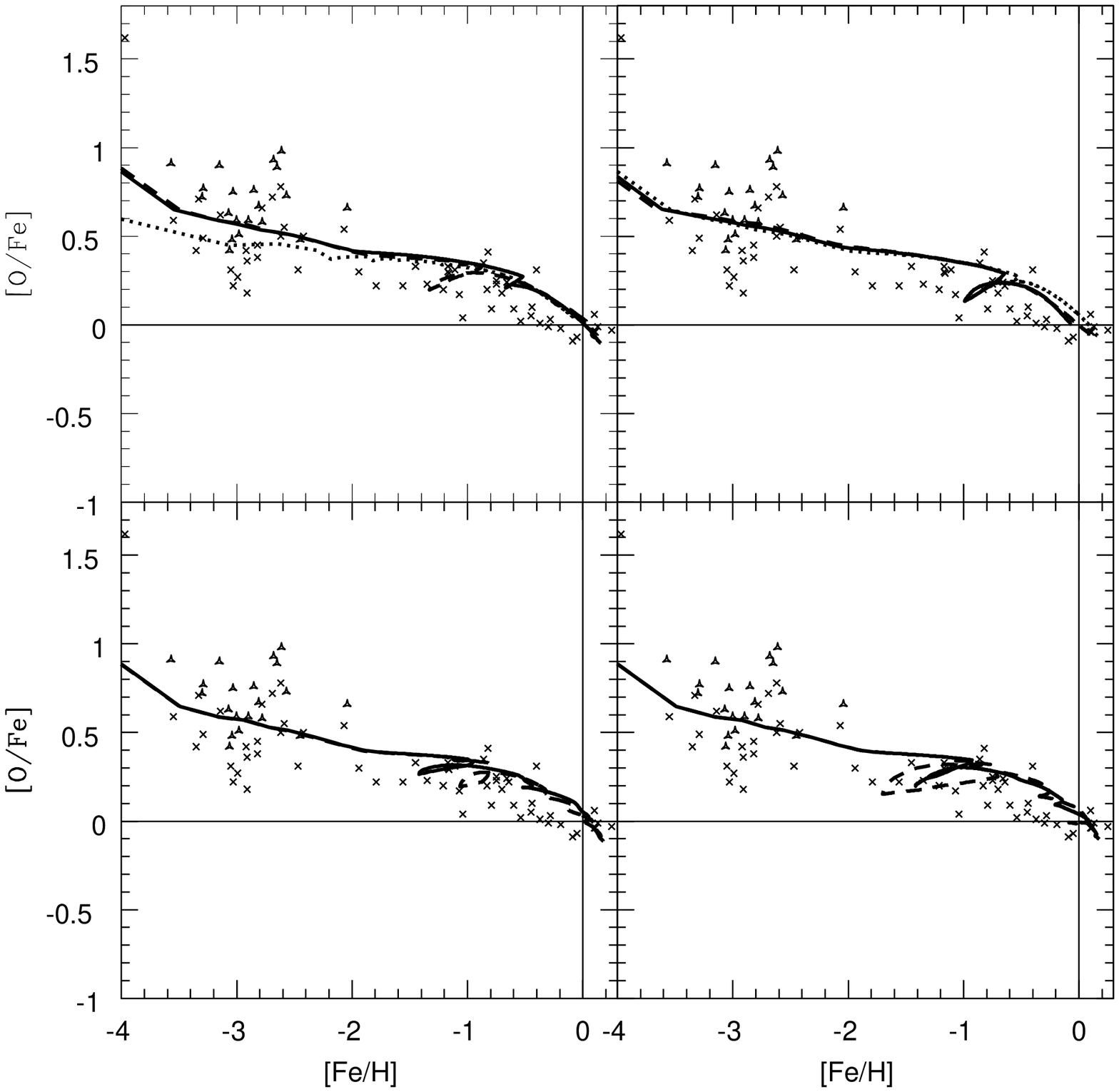}
\caption{[O/Fe] vs [Fe/H]. Upper left panel: red solid line is the two-infall model (Model 1); black dashed line is the cosmological mean model (Model 9); green dotted line is the model by Naab \& Ostriker (2006) (Model 10). Upper right panel: magenta solid line is the constant infall model (Model 2); blue dashed line is the linear infall model (Model 3); cyan dotted line is the pre-enriched model ($Z_{inf} = 1/10 \; Z_{today}$, Model 4). Bottom left panel: black solid line is Model 5; magenta dashed line is Model 6. Bottom right panel: blue solid line is Model 7; cyan dashed line is Model 8. The data are from: Cayrel et al. (2004) (red triangles) and Fran\c cois et al. (2004) (green crosses).}
\end{figure*} 

Figures 12 and 13 present the [Mg/Fe] and the [Si/Fe] as a function of [Fe/H]. The data in figures 10, 11, 12 and 13 are from Cayrel et al. (2004) for the very metal poor stars and from the compilation of Fran\c cois et al. (2004) for all the others. Once again all the considerations made above for [O/Fe] are valid for these other $\alpha$-elements.

\begin{figure*}
\centering
\includegraphics[width=0.8\textwidth]{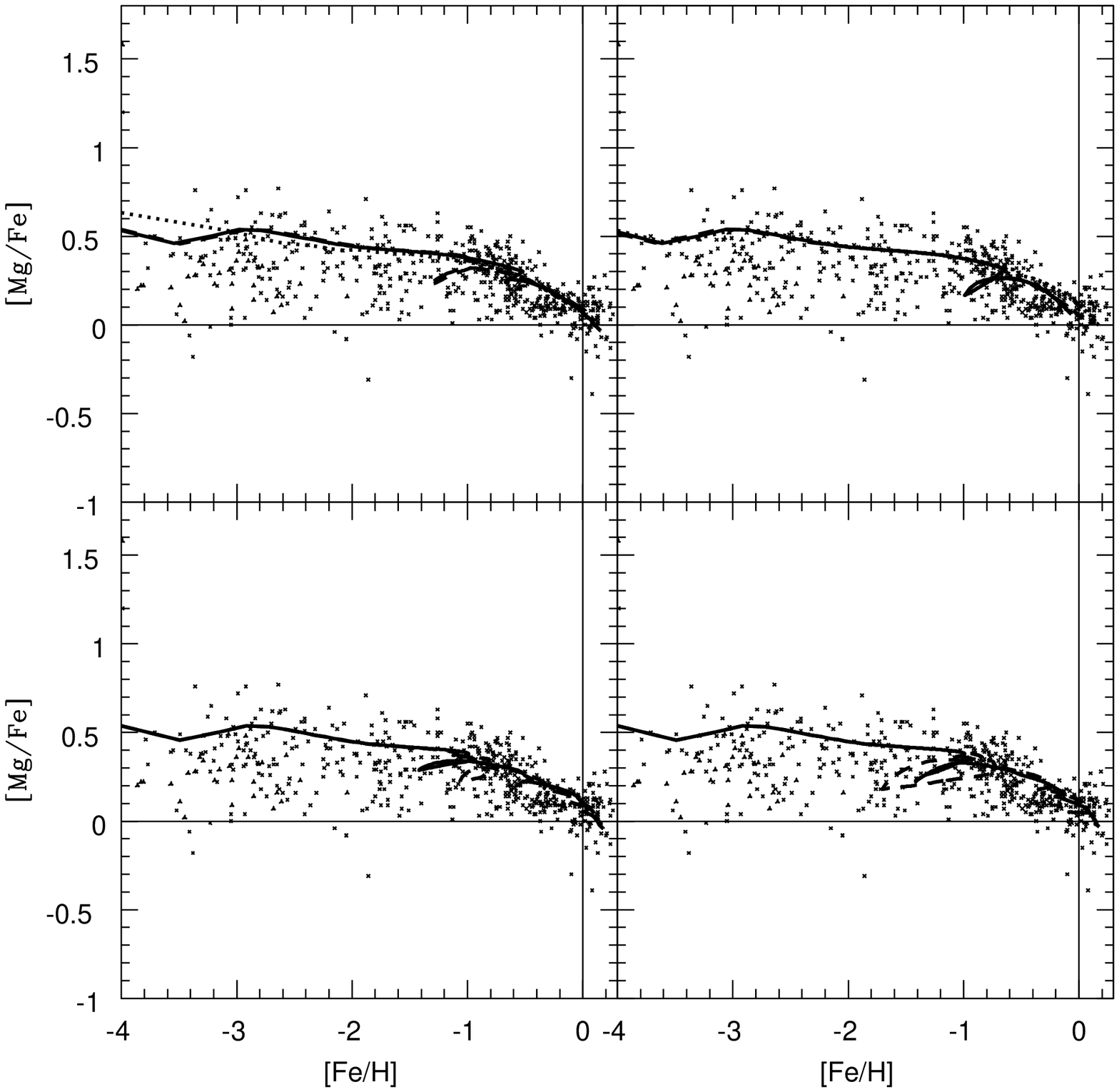}
\caption{[Mg/Fe] vs [Fe/H]. Upper left panel: red solid line is the two-infall model (Model 1); black dashed line is the cosmological mean model (Model 9); green dotted line is the model by Naab \& Ostriker (2006) (Model 10). Upper right panel: magenta solid line is the constant infall model (Model 2); blue dashed line is the linear infall model (Model 3); cyan dotted line is the pre-enriched model ($Z_{inf} = 1/10 \; Z_{today}$, Model 4). Bottom left panel: black solid line is Model 5; magenta dashed line is Model 6. Bottom right panel: blue solid line is Model 7; cyan dashed line is Model 8. The data are from: Cayrel et al. (2004) (red triangles) and Fran\c cois et al. (2004).}
\end{figure*}

\begin{figure*}
\centering
\includegraphics[width=0.8\textwidth]{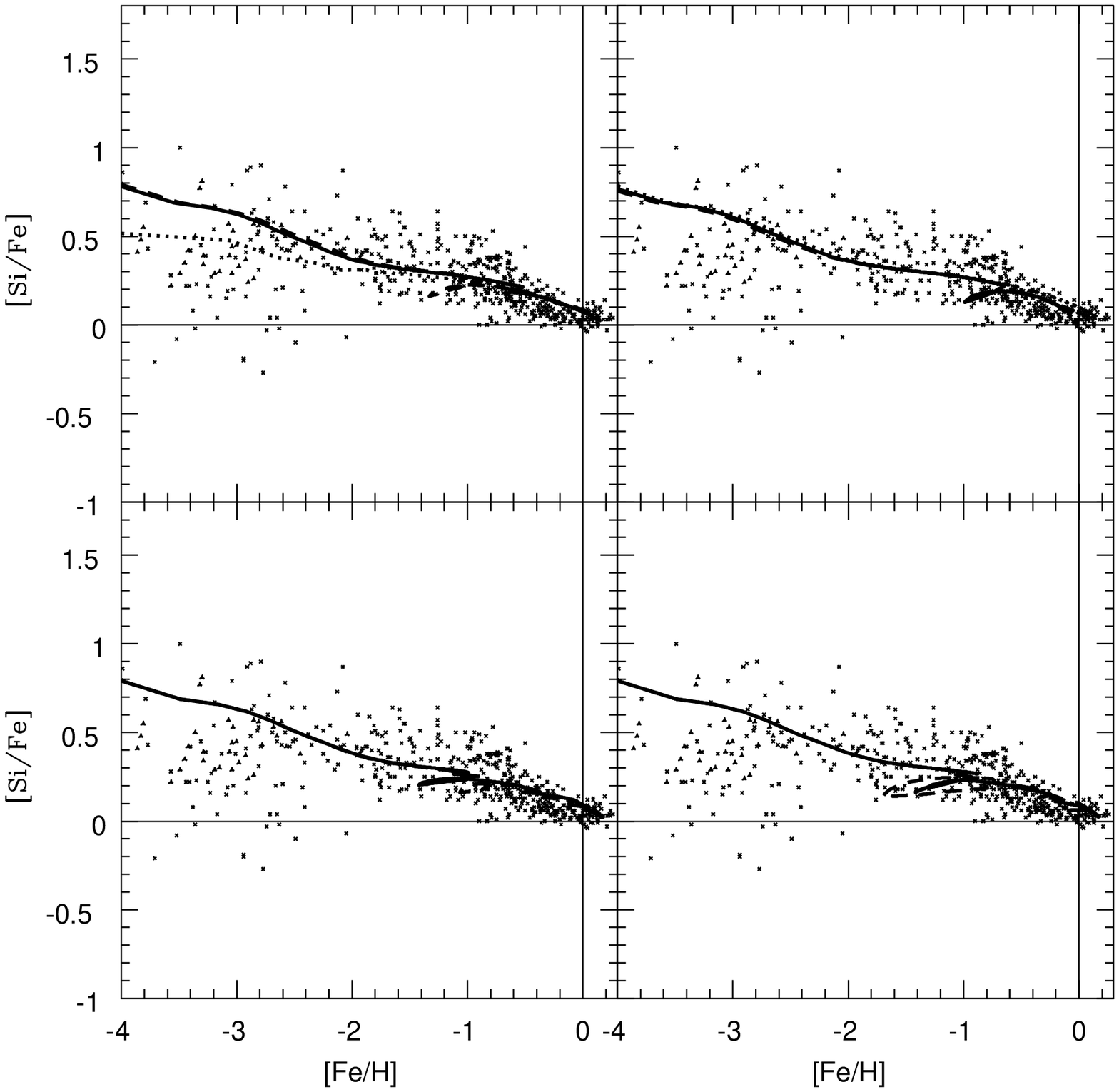}
\caption{[Si/Fe] vs [Fe/H]. Upper left panel: red solid line is the two-infall model (Model 1); black dashed line is the cosmological mean model (Model 9); green dotted line is the model by Naab \& Ostriker (2006) (Model 10). Upper right panel: magenta solid line is the constant infall model (Model 2); blue dashed line is the linear infall model (Model 3); cyan dotted line is the pre-enriched model ($Z_{inf} = 1/10 \; Z_{today}$, Model 4). Bottom left panel: black solid line is Model 5; magenta dashed line is Model 6. Bottom right panel: blue solid line is Model 7; cyan dashed line is Model 8. The data are from: Cayrel et al. (2004) (red triangles) and Fran\c cois et al. (2004).}
\end{figure*} 

Other two important elements are C and N. Figures 14 and 15 show the
behaviour of [C/Fe] and [N/Fe] as a function of [Fe/H]. The data in
figure 14 are from Spite et al. (2005) (magenta points), Carbon et
al. (1987) (red points), Clegg, Lambert \& Tomkin (1981) (cyan
points), Laird (1985) (black points) and Tomkin et al. (1995) (green
points). Figure 15 presents the data from Spite et al. (2005) (magenta
points), Israelian et al. (2004) (blue points), Carbon et al. (1987)
(red points), Clegg, Lambert \& Tomkin (1981) (cyan points) and Laird
(1985) (black points). From  figure 14 it can be seen once again that
the cosmological models are very similar to the model by Chiappini et
al. (1997). 
The predicted curves are different only for values of
[Fe/H] higher than $-1.5$ dex.  The same thing happens for the
[N/Fe]. In both cases, cosmological models have a particular behaviour
at high metallicities. This behaviour is common to all the elements analysed and
is due to the gap in the SFR at about 1 Gyr, as discussed before. In
the cosmological models this effect is larger because of the longer
duration of the gap. However in the case of [C/Fe] and [N/Fe] we
cannot draw any firm conclusion because of the large spread in the
data.
Finally, in figure 16 we show the O abundance gradient as predicted by Model 1 and Model 5, compared with a compilation of data including Cepheids (see Cescutti et al. 2007). As one can see, the O gradient predicted by Model 5 flattens for $r < 8$ kpc whereas agrees very well with the slope predicted by Model 1 (the original two-infall model) for $r  \ge 8$ kpc. Model 1 contains the assumption of an inside-out formation of the disk, as described by eq. (3), whereas in Model 5 no such assumption is made. In spite of that, the two predicted gradients are similar and  we cannot reject the O gradient predicted by Model 5 on the basis of the comparison with data. The reason for that probably resides in the adoption of the star formation threshold which acts mainly at large galactocentric distances where the gas density is lower. This effect is predominating over the increase of the timescale for disk formation. This deserves a more detailed study which we plan to do in a more detail the disk evolution  in a cosmological context in a forthcoming paper.

\begin{figure*}
\centering
\includegraphics[width=0.8\textwidth]{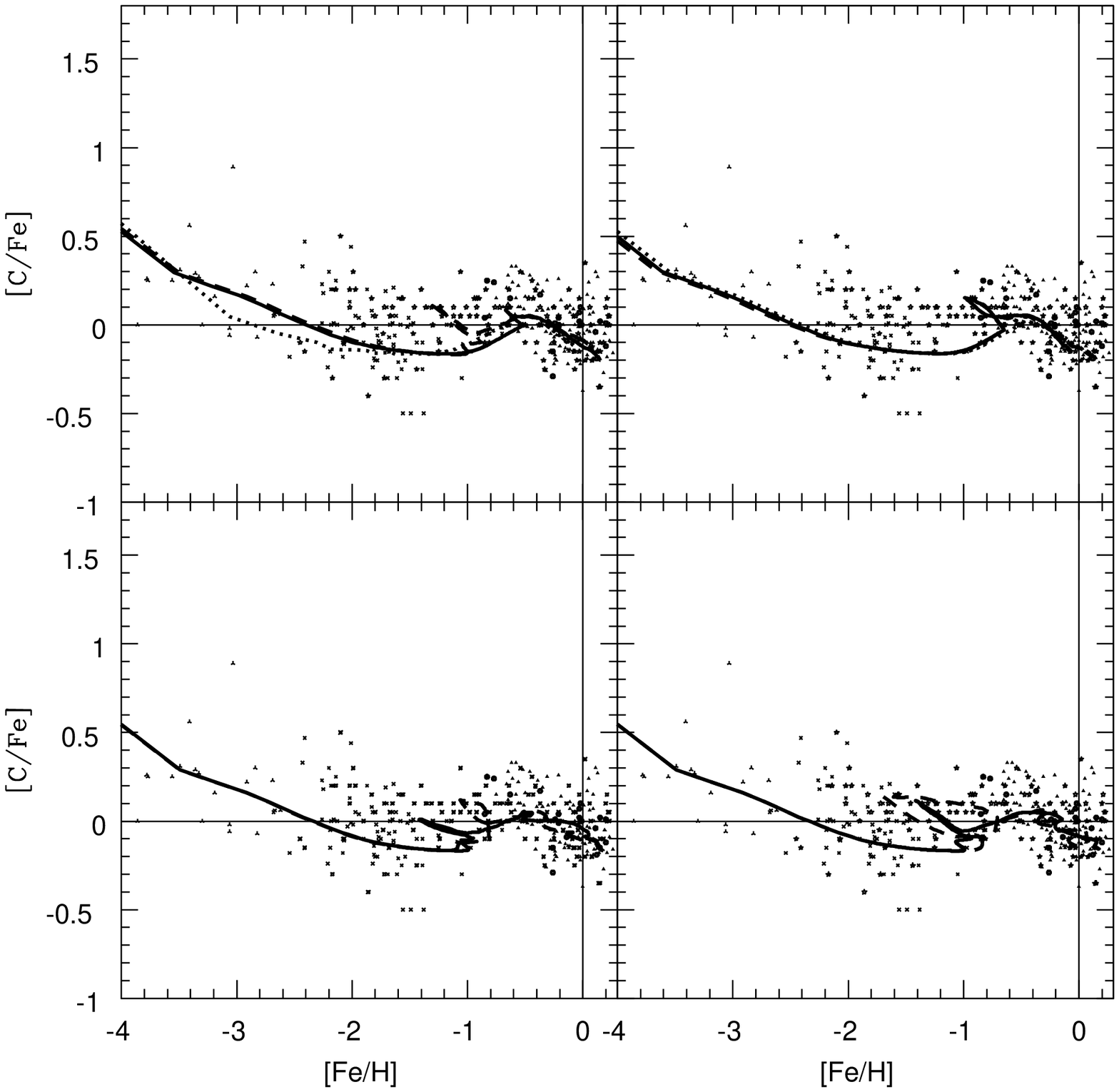}
\caption{[C/Fe] vs [Fe/H]. Upper left panel: red solid line is the two-infall model (Model 1); black dashed line is the cosmological mean model (Model 9); green dotted line is the model by Naab \& Ostriker (2006) (Model 10). Upper right panel: magenta solid line is the constant infall model (Model 2); blue dashed line is the linear infall model (Model 3); cyan dotted line is the pre-enriched model ($Z_{inf} = 1/10 \; Z_{today}$, Model 4). Bottom left panel: black solid line is Model 5; magenta dashed line is Model 6. Bottom right panel: blue solid line is Model 7; cyan dashed line is Model 8. The data are from: Spite et al. (2005) (magenta stars with three arms), Carbon et al. (1987) (red crosses), Clegg, Lambert \& Tomkin (1981) (cyan circles), Laird (1985) (black stars with five arms) and Tomkin et al. (1995) (green triangles).}
\end{figure*} 

\begin{figure*}
\centering
\includegraphics[width=0.8\textwidth]{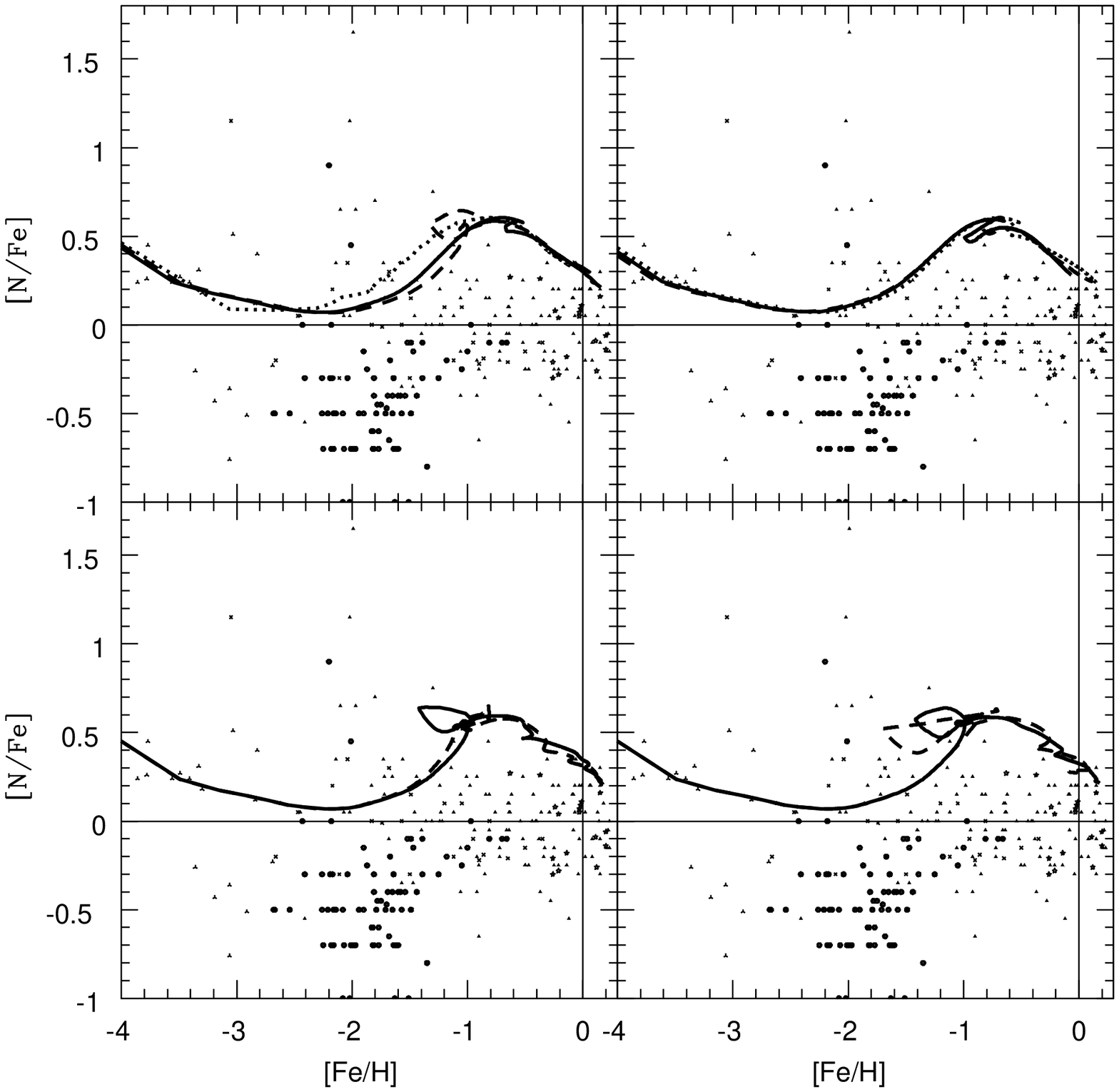}
\caption{[N/Fe] vs [Fe/H]. Upper left panel: red solid line is the two-infall model (Model 1); black dashed line is the cosmological mean model (Model 9); green dotted line is the model by Naab \& Ostriker (2006) (Model 10). Upper right panel: magenta solid line is the constant infall model (Model 2); blue dashed line is the linear infall model (Model 3); cyan dotted line is the pre-enriched model ($Z_{inf} = 1/10 \; Z_{today}$, Model 4). Bottom left panel: black solid line is Model 5; magenta dashed line is Model 6. Bottom right panel: blue solid line is Model 7; cyan dashed line is Model 8. The data are from: Spite et al. (2005) (magenta stars with three arms), Israelian et al. (2004) (blue crosses), Carbon et al. (1987) (red circles), Clegg, Lambert \& Tomkin (1981) (cyan stars with five arms) and Laird (1985) (black triangles).}
\end{figure*}

\begin{figure*}
\centering
\includegraphics[width=0.8\textwidth]{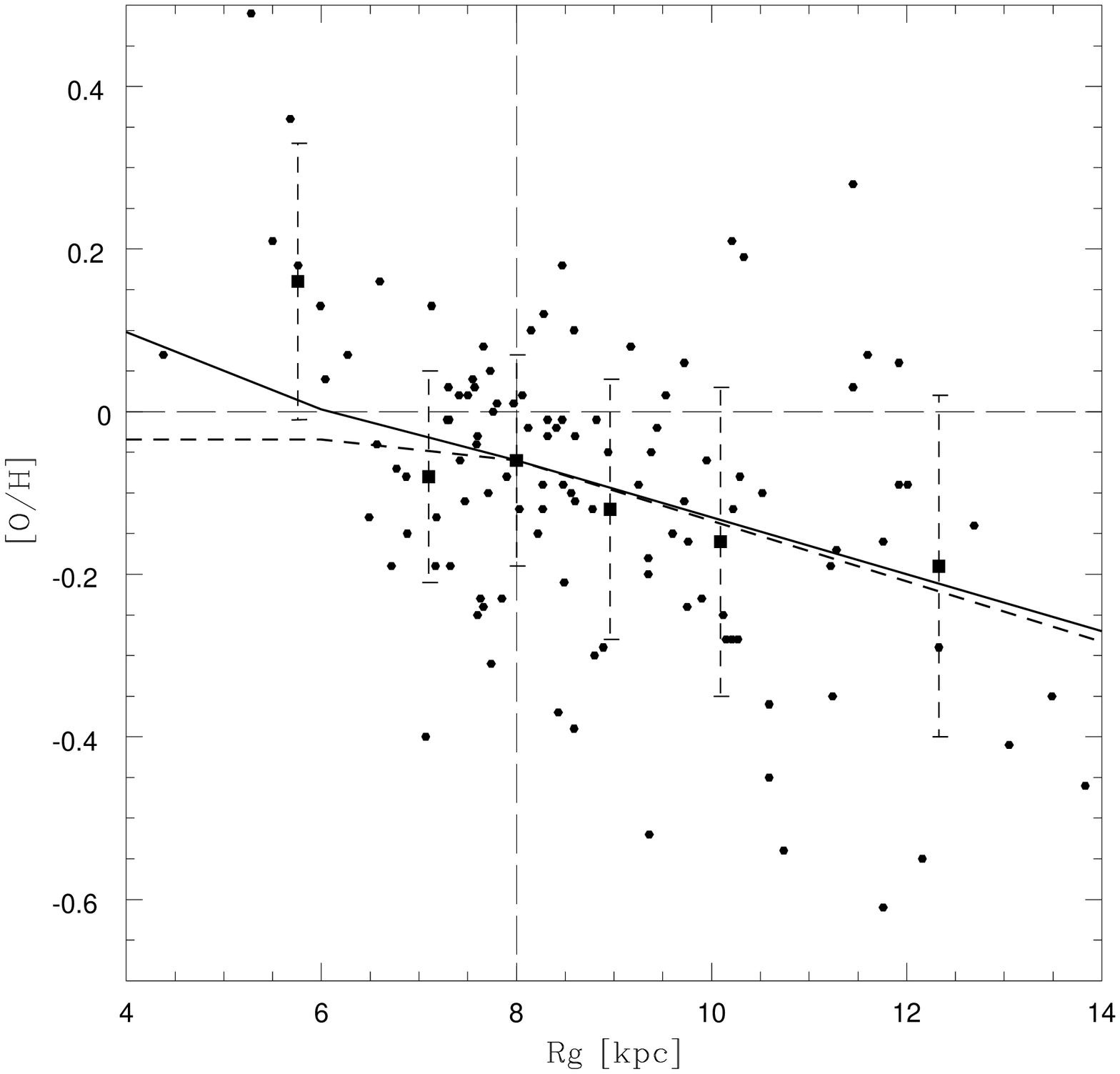}
\caption{Predicted and observed O abundance gradients in the galactocentric 
distance range 4 - 14 kpc. The continuous line is the prediction of the two-infall model, whereas the dashed line is the prediction of Model 5. The data points are from Cepheids. The big squares with error bars represent averages of the points with their errors (see Cescutti et al. 2007 and reference therein).}
\end{figure*} 
  
Figures 17, 18 and 19 present the results obtained by using a
different infall law, derived from the cosmological simulation but
selecting different parameters. In this case we selected a halo which
is not expected to produce a spiral galaxy, so we looked for a spin parameter
lower than 0.04, a redshift of last major merger lower than 2.5 and a
redshift of formation lower than 1.0. Such a halo is perhaps more
appropriate for an elliptical or S0 galaxy. We
found a halo with the following characteristics:
\begin{itemize}
\item mass = $2.15 \cdot 10^{12} M_{\odot}$
\item $\lambda = 0.029$
\item redshift of major merger = 0.50
\item redshift of  formation = 0.75 - 0.63  
\end{itemize} 
Figure 17, 18 and 19 compare the results from this halo with Model 1
(two-infall law) and Model 5 (our best cosmological choice). The
infall law is very different. In particular, it has a major peak at a redshift of about
0.3. This produces a peak at the same redshift in the star formation
rate and, of course, in the SNII rate. Moreover, there is a strong
depression in the [Fe/H] ratio between 1.8 and 3 Gyr from the
beginning of the simulation, difficult to reconcile with observations.

In figures 18 and 19 we show the results for the [O/Fe] and for the
G-dwarf metallicity distribution. The
main difference between this halo and Models 1 and 5 is that the loop
placed at [Fe/H] $\sim -1.0$ is longer and predicts low values of
[O/Fe] at low [Fe/H], which is not observed in Galactic stars. As far as the G-dwarf
metallicity distribution is concerned, the halo forms too many stars with low metallicity as a consequence of the deep
depression in the [Fe/H] ratio (see the plot on the bottom right part of figure 17), again not in agreement with the data,
and resembles an early-type galaxy.
This example confirms the importance of the cosmological assembly
history of the DM halo in determining not only the morphological
parameters of the galaxy it hosts, but also its chemical properties.

\begin{figure*}
\centering
\includegraphics[width=0.4\textwidth]{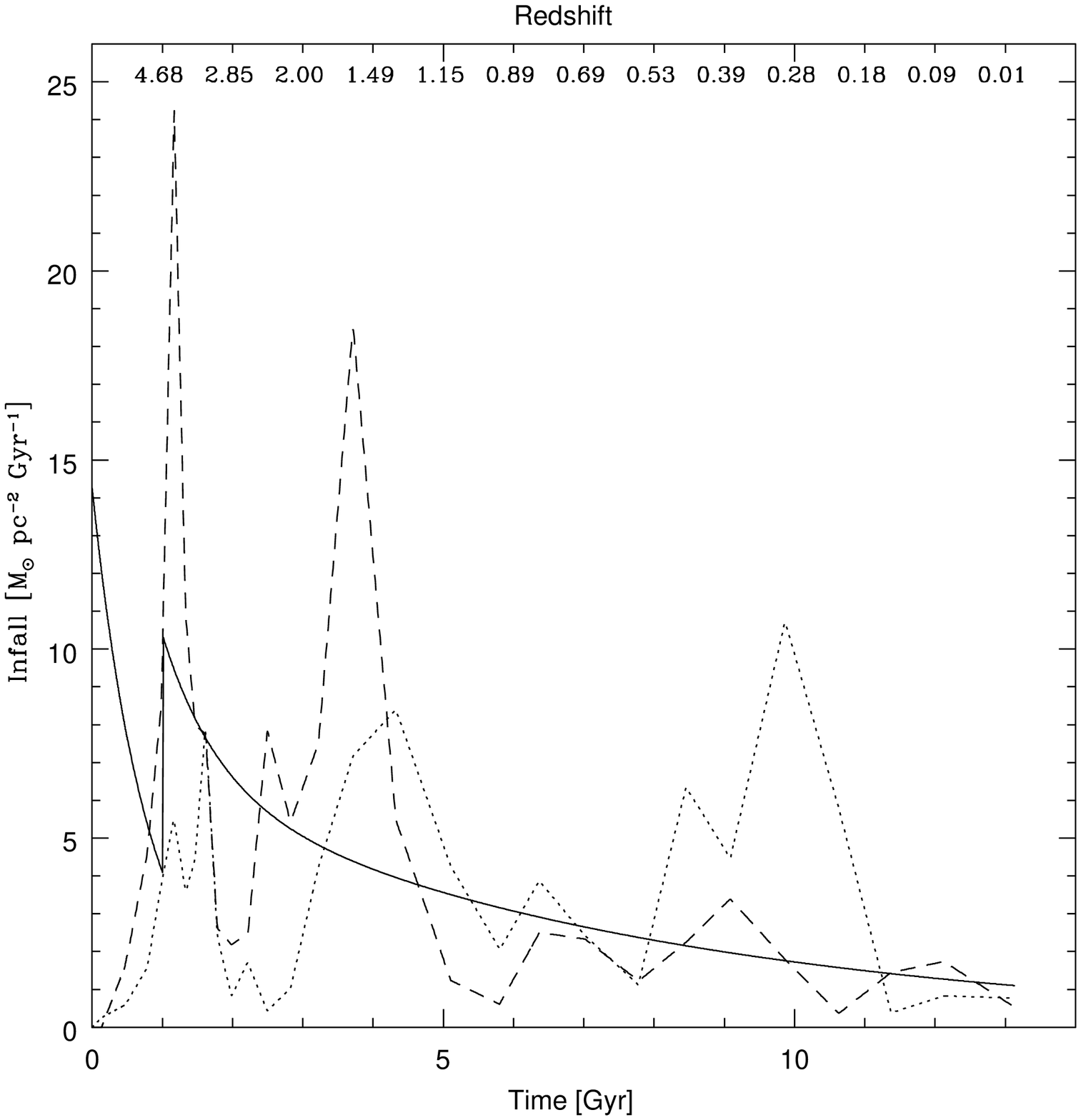}
\includegraphics[width=0.4\textwidth]{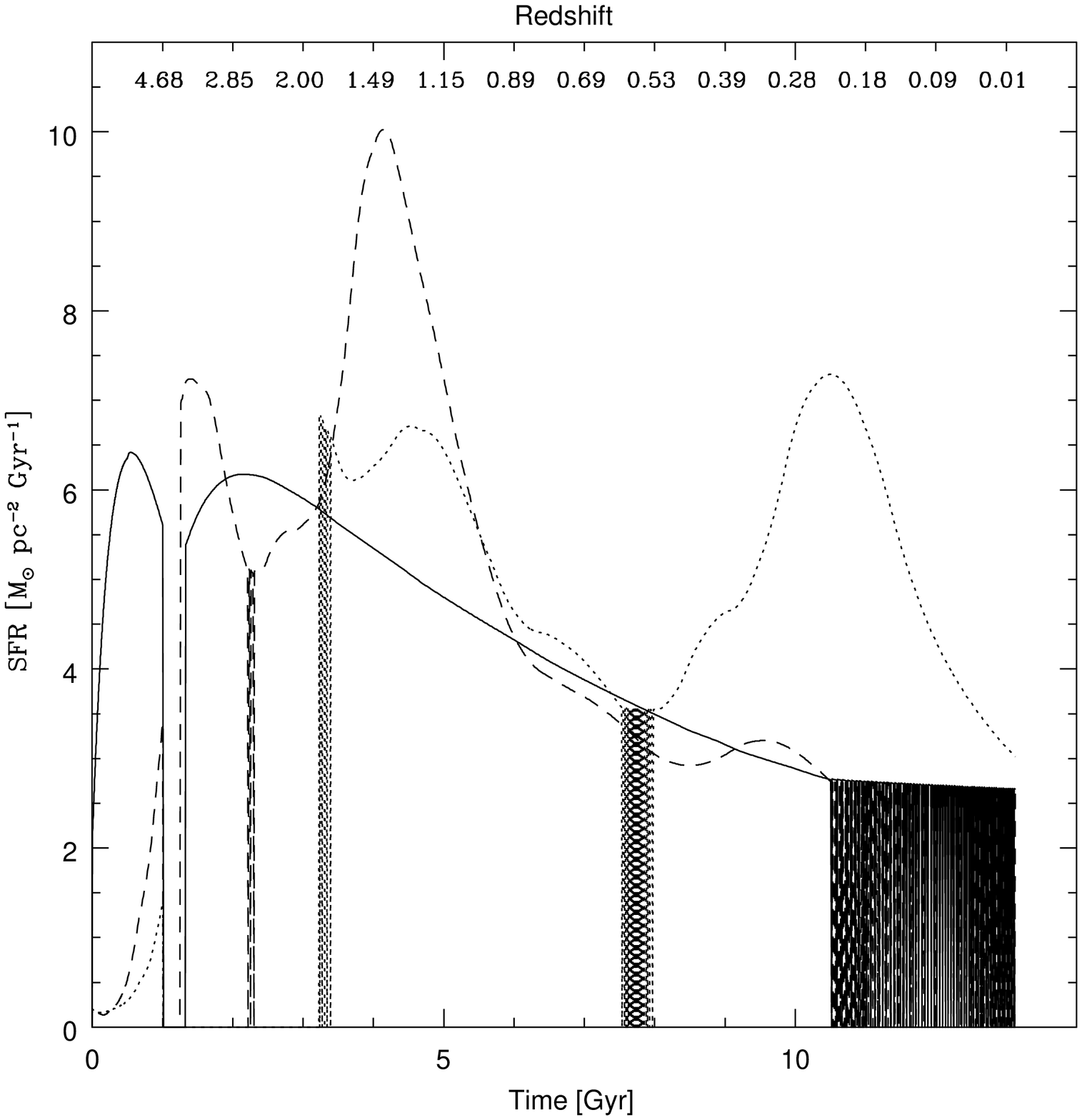}

\includegraphics[width=0.4\textwidth]{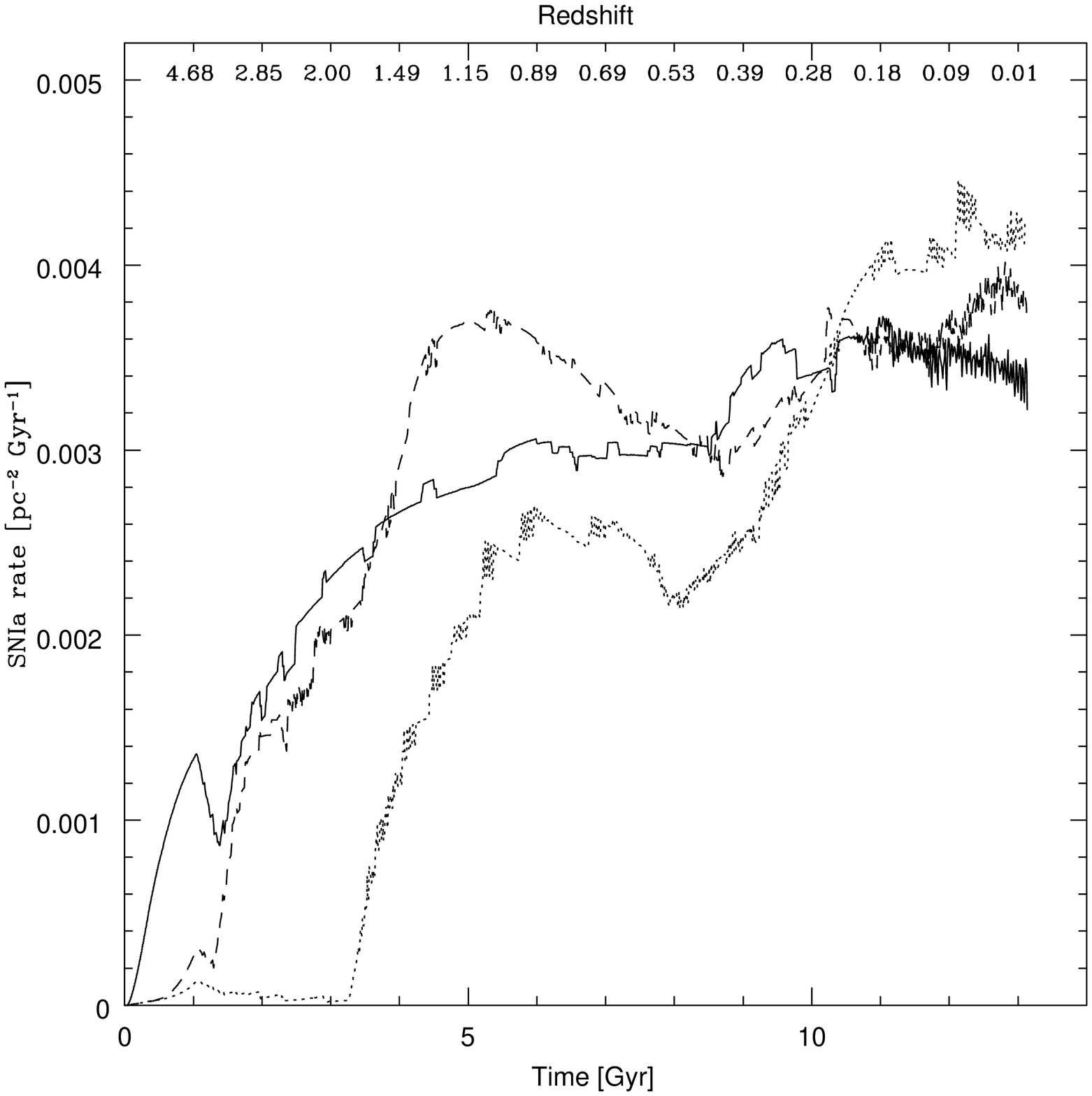}
\includegraphics[width=0.4\textwidth]{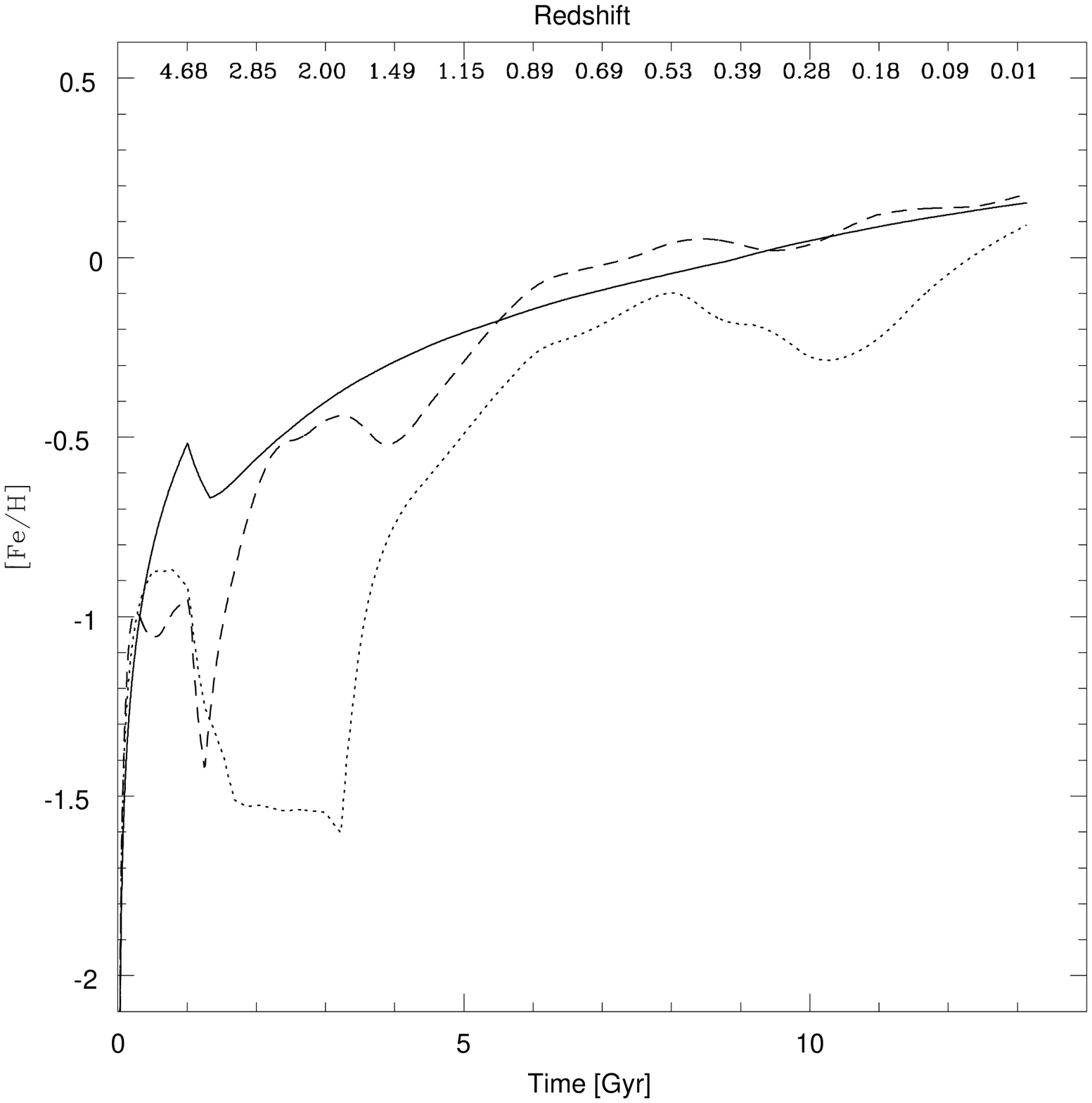}
\caption{These plots represent the infall law (upper left panel), the star formation rate (upper right panel), the SNIa rate (bottom left panel) and the [Fe/H] (bottom right panel) as a function of time for the two-infall model (Model 1, red solid line), for Model 5 (black dashed line) and for the halo 20912 (blue dotted line).}
\end{figure*} 

\begin{figure*}
\centering
\includegraphics[width=0.8\textwidth]{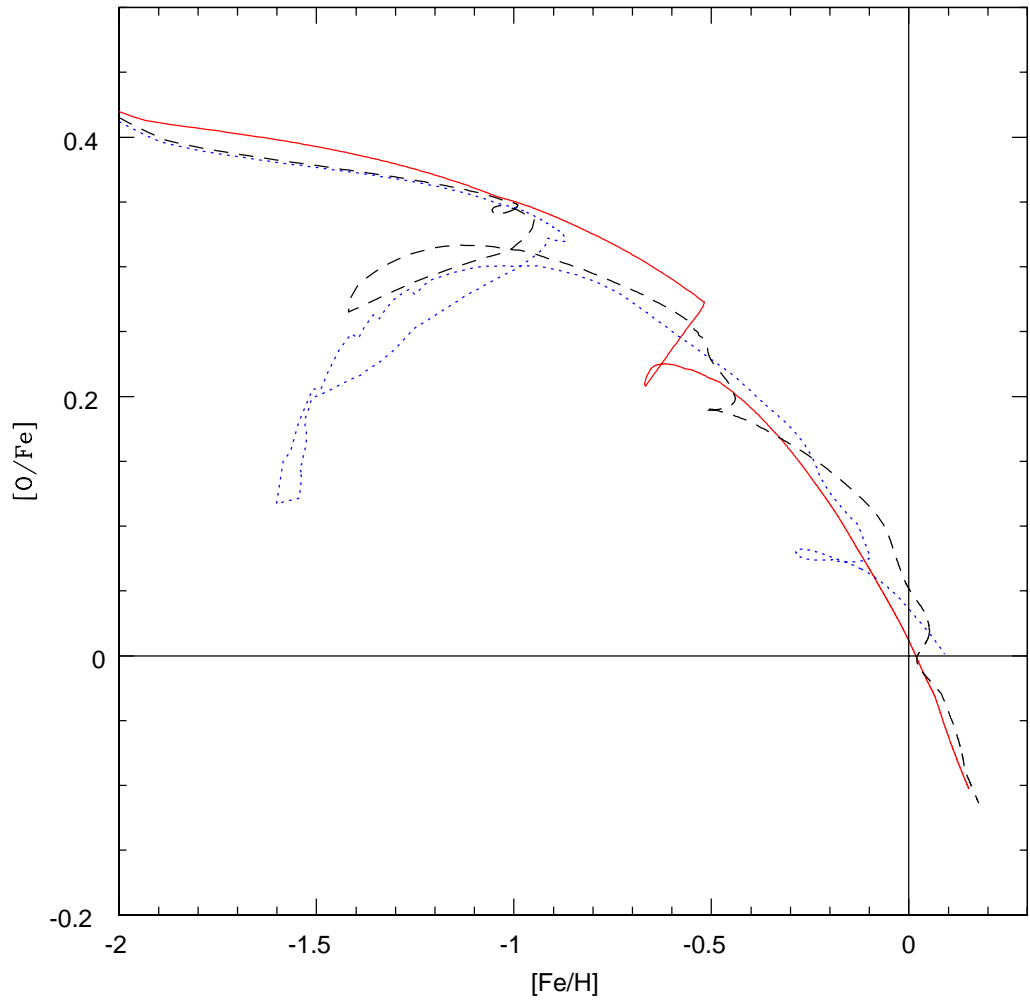}
\caption{This plot represents the [O/Fe] as a function of [Fe/H]. The red solid line represents the two-infall model (Model 1), the black dashed line represents Model 5 and the blue dashed line the halo 20912.}
\end{figure*} 

\begin{figure*}
\centering
\includegraphics[width=0.8\textwidth]{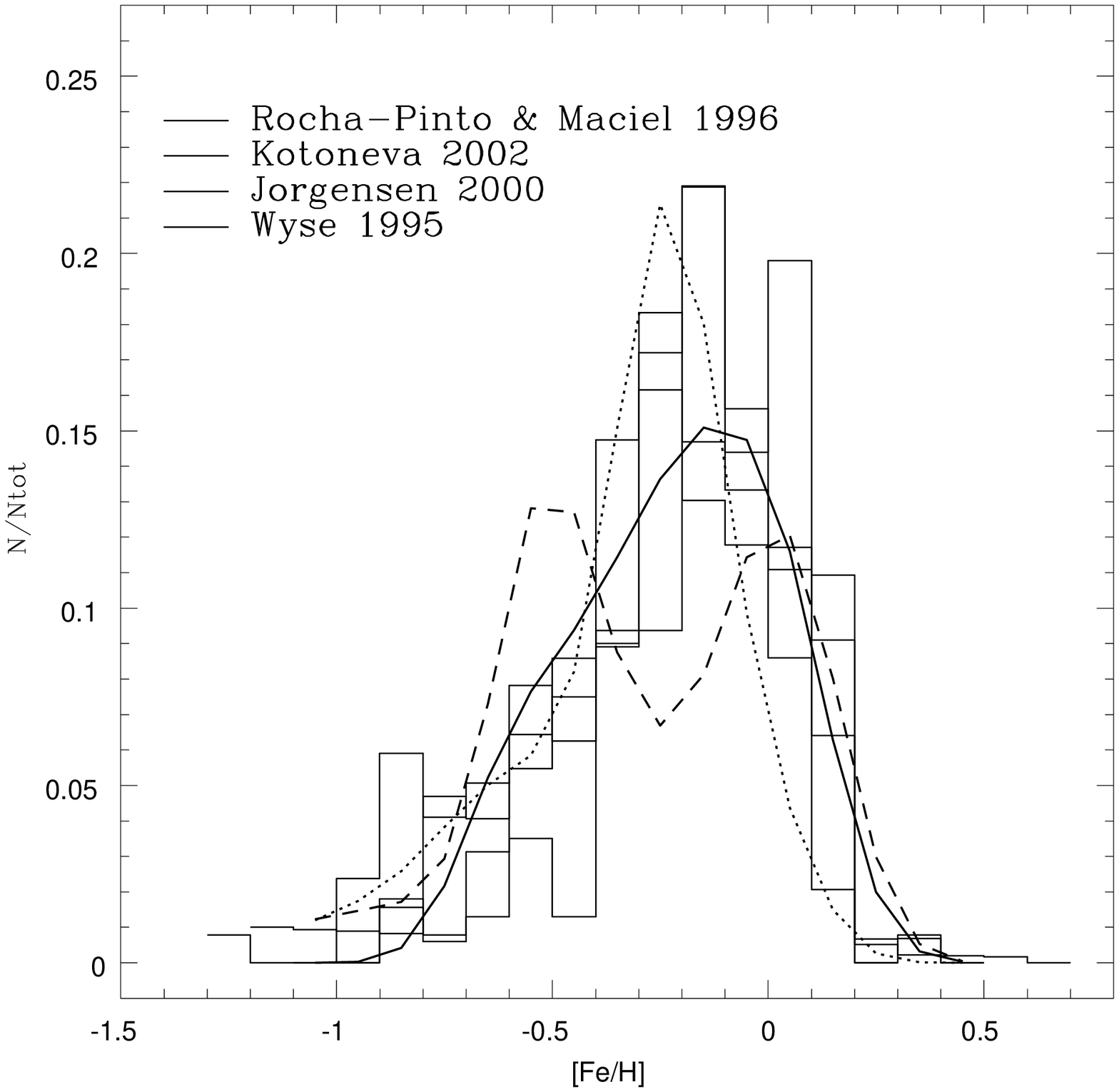}
\caption{G-dwarf metallicity distribution for the two-infall model (red solid line), Model 5 (black dashed line) and for the halo 20912 (blue dashed line).}
\end{figure*}

\section{Conclusions}
We have tested different gas infall laws for models of the formation of the
Milky Way and especially cosmologically derived infall laws, obtained
by means of cosmological simulations for the formation of the DM halo
of the Milky Way. In particular, we assumed that the accretion law for
the DM halo holds also for the baryonic matter. We
found four different DM halos with properties compatible with a disk
galaxy, with one in particular seeming better than the others.  All
these infall laws were then compared with the one proposed by
Chiappini et al. (1997), called two-infall law, which predicts that
there were two main accretion episodes which formed the halo-bulge-thick
disk and the thin disk, respectively. We found that our best
cosmological infall law is very similar to the two-infall one, which
has already proven to be able to reproduce the majority of the
chemical properties of the Milky Way in the solar neighbourhood.
Our cosmological infall laws have been tested in a detailed chemical
evolution model for the Milky Way, following the evolution of several
chemical elements by taking into account stellar lifetimes, SN
progenitors and stellar nucleosynthesis.

Our main conclusions can be summarized as follows:
\begin{itemize}
\item 

A model with constant infall predicts a present day infall rate and
SFR larger than all the other models. Moreover, it is the only
model which produces an unrealistically increasing SFR during the last
billion years. This is probably an unrealistic law, and we
only used  for a purpose of comparison with other
infall laws.

\item The linear model predicts the largest amount of stars presently in the solar neighbourhood but it  seems to reproduce reasonably well all the other observables. However, this model does not describe the evolution of our Galaxy as well as an exponential law does.

\item  The model adopting the two-infall law but where the gas is assumed to be pre-enriched during the formation of the disk at the level of 1/10 of solar well reproduces the G-dwarf metallicity distribution, as expected.

\item The cosmological laws, and in particular our preferred best fit, seem to fit
well all the data. This law predicts two main accretion episodes which
can be identified with the formation of halo-thick disk and thin disk
, respectively, very similar to the two-infall law. Moreover, there
seems to be a gap of 1-2 Gyr in the SFR between the two episodes,
larger than predicted by Chiappini et al. (1997) ($<$ 1 Gyr).  The gap
is due mainly to the adoption of a threshold gas density for the star
formation rate. Such a gap seems to have been observed looking at
abundance patterns, in particular at [Fe/O] vs. [O/H] (Gratton et
al. 1996) and at [Fe/Mg] vs. [Fe/H] (Fuhrmann 1998), although new data
are necessary to draw firm conclusions. The model including this
cosmological infall law can well reproduce most of the observational
constraints. It predicts for the G-dwarf metallicity distribution, in the solar vicinity, two
different peaks: we speculate that the first peak represents the stars 
of the halo and thick disk while the second peak represents the stars of the thin disk. The same metallicity distribution computed for the central region should include also the bulge stars. The
predicted timescales for the formation of the halo-thick disk and the
thin disk, respectively, are in excellent agreement with those
suggested by Chiappini et al.  In particular, the halo-thick disk
must have formed on a timescale not longer than 1-2 Gyr whereas the
thin disk in the solar vicinity took at least 6 Gyr to assemble 60\% of its mass.  As a
consequence of the gap between the halo-thick disk and the thin disk, we 
predict that the thin disk is at least 2 Gyr younger than the halo.

\item The other cosmological infall laws are characterized by several
minor accretion events after the two main ones and predict larger
gaps in the SFR which are not observed in the [Fe/O] vs. [O/H] and [Fe/Mg] vs. [Mg/H] which indicate a gap not larger than 1-2 Gyr.

\item A model adopting a cosmologically inferred infall law by  Naab \& Ostriker (2006) presents a behaviour very similar to the constant infall law and predicts too low metallicities at the Sun age and at the present time. Moreover, this model predicts a too small number of G-dwarf with high metallicity. In their paper they present a G-dwarf metallicity distribution but as a function of Z which represents O and not Fe as in the observations. 

\item Our results strongly depend on what criteria were used to select the dark matter halo from the cosmological simulations. If they are not suitable for forming a spiral galaxy it is
possible to see that the results are not in good agreement with the
observations. We prove it by using a DM halo with dynamical parameters
compatible with an early-type galaxy.

\item Our results can be compared with the work of Robertson et
 al. (2005), in which the authors studied the chemical enrichment of
 the stellar halo of the Milky Way, using the prescriptions of the
 hierarchical scenario. They supposed that most of the mass in the MW
 halo was acquired via mergers with massive dIrr-type DM halos,
 occurred at a look-back time of $\sim$ 10 Gyr. They used three
 examples of mass accretion history, supposing that the cumulative
 mass accretion in individual DM halos can be well described by a
an analytical function obtained by Wechsler et al (2002).

Moreover, they assumed that
the cold gas inflow rate tracks the DM accretion rate and that the
fraction of cold gas is equal to 2\%.

In order to build the stellar halo of the Milky Way they used a
dIrr-type dark matter halo with a virial mass $M_{0} = 6$ x $10^{10}
M_{\odot}$, accreted 9 Gyr ago, following their assumed
accretion law. In this case the time available for the star formation and 
the consequent chemical enrichment is only $\sim$ 2.6 Gyr and therefore 
the chemical enrichment due to SNIa was limited.
We do not use the accretion of a dIrr galaxy to build the
stellar halo of the MW. We obtained the mass accretion history of the DM
halo directly from the cosmological simulation, done with the public
tree-code GADGET2 (Springel 2005). For this reason we accrete only DM
and cold gas and not already formed dwarf galaxies, with their own stars
and gas. Moreover we study the chemical enrichment of all the galaxy and
not only of the stellar halo.

\item In the future we plan to extend the current work, and in particular our
cosmologically derived baryonic infall laws, to the study of the chemical
properties of the whole disk. As we have already shown in this paper, by normalizing the infall law to the  present time total surface mass density along the disk, we obtain different timescales for the assembly of the disk as a function of galactocentric distance, although the inside-out effect is not as marked as in the Matteucci \& Fran\c cois (1989) and Chiappini et al. (2001) models.

\end{itemize}

The fact that all our four suitable DM halos show an accretion law which
resembles that used in the two-infall model could be linked to the way in
which such halos assemble. Indeed, they have their last major merger at high
redshift, larger than $z=2.5$, by selection and they reach a mass
larger than 50\% of their final one at lower redshift. As a consequence of our
requirement not to have late major mergers, such late assembly happens via
accretion of material from the field, namely filaments, or via minor mergers.
These two epochs of important accretion qualitatively corresponds to the two
peaks used in the two-infall model and give it a cosmological motivation.
Obviously the details of the late accretion episode will depend on the
dynamical history of the single DM halo, and will generate differences in the
chemical patterns of individual late-type galaxies without destroying their
overall properties.  
On the other hand, halos should acquire their angular momentum thanks to
the cosmological torques acting at high redshifts on the material (both
baryons and dark matter) which will coalesce to form them. Such torques will
also influence their mass accretion histories.  Thus, selecting DM halos with
high spin values could also result in selecting halos with similar dynamical
histories. Astrophysical processes acting on baryons, e.g. feedback, should
not be able to dramatically alter this scenario.
Finally we note that, while in the two-infall model the timing of the
two episodes is a free parameter, in the cosmological infall scenario
the timing is directly given by the gravitational evolution of the
halos. In this sense, the agreement between such models is not a-priori
guaranteed and could be interpreted as an interesting link between the
morphological properties of the late-type galaxies (used to fix our
requirement) and their chemical properties, via the hierarchical model.

\section{Acknowledgments}

We thank Gabriele Cescutti and Cristina Chiappini for some helpful suggestions and Donatella Romano, Patrick Fran\c cois and Silvia Kuna Ballero for their collaboration concerning the data. We also thank the referee Chris Flynn for valuable comments. 
\vfill\eject

\label{lastpage}

\end{document}